\documentclass[12pt]{article}
\usepackage{amsmath,color}
\usepackage[dvips]{graphicx}
\input amssym
\input amssym.def

\def\red#1{{\color{red} #1}}

\begin{document}

\def\prg#1{\par\medskip\noindent{\bf #1}}  \def\ra{\rightarrow}
\newcounter{nbr}
\def\note#1{\bitem\vspace{-5pt}\addtocounter{nbr}{1}
            \item{} #1\vspace{-5pt}
            \eitem}
\def\lra{\leftrightarrow}              \def\Lra{{\Leftrightarrow}} \def\Ra{\Rightarrow}                   \def\La{\Leftarrow}
\def\nin{\noindent}                    \def\pd{\partial}
\def\dis{\displaystyle}
\def\grl{{GR$_\Lambda$}}               \def\vsm{\vspace{-8pt}}
\def\Leff{\hbox{$\mit\L_{\hspace{.6pt}\rm eff}\,$}}
\def\bull{\raise.25ex\hbox{\vrule height.8ex width.8ex}}
\def\ric{{Ric}}                        \def\tmgl{\hbox{TMG$_\Lambda$}}
\def\Lie{{\cal L}\hspace{-.7em}\raise.25ex\hbox{--}\hspace{.2em}}
\def\hd{{^\star}}                      \def\dis{\displaystyle}
\def\mb#1{\hbox{{\boldmath $#1$}}}     \def\phb{\phantom{\Big|}}
\def\ph#1{\phantom{#1}}                \def\diag{\text{diag}}
\def\tr{\text{tr}\,{}}                 \def\det{\,\text{det}{}}
\def\PG+{{\text{PG}$^+$}}              \def\PGm{{\text{PG}$^-$}}

\def\hook{\hbox{\vrule height0pt width4pt depth0.3pt
\vrule height7pt width0.3pt depth0.3pt
\vrule height0pt width2pt depth0pt}\hspace{0.8pt}}
\def\inn{\hook}
\def\first{\rm (1ST)}  \def\second{\hspace{-1cm}\rm (2ND)}

\def\G{\Gamma}        \def\S{\Sigma}        \def\L{{\mit\Lambda}}
\def\D{\Delta}        \def\Th{\Theta}       \def\Ups{\Upsilon}
\def\a{\alpha}        \def\b{\beta}         \def\g{\gamma}
\def\d{\delta}        \def\m{\mu}           \def\n{\nu}
\def\th{\theta}       \def\k{\kappa}        \def\l{\lambda}
\def\vphi{\varphi}    \def\ve{\varepsilon}  \def\p{\pi}
\def\r{\rho}          \def\Om{\Omega}       \def\om{\omega}
\def\s{\sigma}        \def\t{\tau}          \def\eps{\epsilon}
\def\nab{\nabla}      \def\btz{{\rm BTZ}}   \def\heps{\,{\hat\eps}}

\def\bR{\bar{R}}       \def\bT{\bar{T}}     \def\cL{{\cal L}}
\def\bcL{\bar{\cL}}   \def\hc{\hat{c}}
\def\tR{{\tilde R}}   \def\tU{{\tilde U}}  \def\cE{{\cal E}}
\def\cH{{\cal H}}     \def\hcH{\hat{\cH}}  \def\cT{{\cal T}}
\def\cK{{\cal K}}     \def\hcK{\hat{\cK}}  \def\cA{{\cal A}}
\def\cV{{\cal V}}     \def\tom{{\tilde\omega}} \def\cR{{\cal R}}
\def\tcL{{\tilde\cL}} \def\bh{{\bar h}}     \def\barf{\bar{f}}
\def\orth{{\perp}}    \def\bm{{\bar m}}     \def\bn{{\bar n}}
\def\bi{{\bar\imath}} \def\bj{{\bar\jmath}} \def\bk{{\bar k}}
\def\bPhi{\bar\Phi}   \def\bW{{\bar W}}     \def\tW{{\widetilde W}}
\def\chm{ \checkmark} \def\bG{{\bar G}}     \def\chmr{\red{\chm}}
\def\bara{{\bar a}}   \def\barb{{\bar b}}   \def\barA{{\bar A}}
\def\barc{{\bar c}}   \def\tU{\tilde{U}}   \def\tPhi{{\tilde\Phi}} \def\tPsi{{\tilde\Psi}} \def\tTh{{\tilde\Th}} \def\tU{{\tilde U}}
\def\bl{{\bar l}}     \def\hs#1{\hspace{#1pt}} \def\what{\widehat}
\def\ul#1{\underline{#1}}                   \def\barB{{\bar B}}
\def\ol#1{{\bar #1}}
\def\nR{{}^{(n)}\hspace{-1pt}R}             \def\nT{{}^{(n)}T}
\def\irr#1#2{\hspace{1pt}{}^{(#1)}\hspace{-2pt}#2{}}
\def\ir#1{\,{}^{#1}\hspace{-1.2pt}}
\let\Pi\varPi       \def\hpi{{\hat\pi}}     \def\hPi{{\hat\Pi}}
\def\hR{{\hat R}}   \def\brK{{\bar K}}      \def\brN{{\bar N}}
\def\brU{{\bar U}}  \def\tW{{\widetilde W}} \def\tN{{\tilde N}}

\vfuzz=2.5pt 
\def\nn{\nonumber}
\def\be{\begin{equation}}             \def\ee{\end{equation}}
\def\ba#1{\begin{array}{#1}}          \def\ea{\end{array}}
\def\bea{\begin{eqnarray}}            \def\eea{\end{eqnarray} }
\def\beann{\begin{eqnarray*}}         \def\eeann{\end{eqnarray*} }
\def\beal{\begin{eqalign}}            \def\eeal{\end{eqalign}}
\def\lab#1{\label{eq:#1}}             \def\eq#1{(\ref{eq:#1})}
\def\bsubeq{\begin{subequations}}     \def\esubeq{\end{subequations}}
\def\bitem{\begin{itemize}}           \def\eitem{\end{itemize}}
\renewcommand{\theequation}{\thesection.\arabic{equation}}

\title{General Poincar\'e gauge theory: Hamiltonian structure\\
       and particle spectrum}
\author{M. Blagojevi\'c and B. Cvetkovi\'c\footnote{
        Email addresses: {\tt mb@ipb.ac.rs, cbranislav@ipb.ac.rs}}\\
Institute of Physics, University of Belgrade \\
                      Pregrevica 118, 11080 Belgrade, Serbia}
\date{\today}                                                
\maketitle

\begin{abstract}
Basic aspects of the Hamiltonian structure of the
parity-violating Poincar\'e gauge theory are studied. We found all
possible primary constraints, identified the corresponding critical
parameters, and constructed the generic form of the canonical Hamiltonian. In addition to being important in their own right, these results offer dynamical information that is essential for a proper understanding of the particle spectrum of the theory, calculated in the weak field approximation around the Minkowski background.
\end{abstract}

\section{Introduction}

Weyl's idea of gauge invariance \cite{x1} turned out to be a key
principle underlying the dynamical structure of all the fundamental physical
interactions. Following this idea and the subsequent works of Yang, Mills
and Utiyama \cite{x2}, Kibble and Sciama \cite{x3} formulated a new theory
of gravity, the Poincar\'e gauge theory (PG, aka PGT), based on gauging
the Poincar\'e group of spacetime symmetries. In PG, spacetime is
characterized by a Riemann-Cartan geometry, in which the torsion and 
curvature are the field strengths associated with the translation and
Lorentz subgroups of the Poincar\'e group; for more details, see
\cite{x4,x5,x6,x7,x8,x9,x10}.

Earlier investigations of PG were mostly focused on the class of
\emph{parity preserving} Lagrangians quadratic in the field strengths;
see, for instance, Hayashi and Shirafuji \cite{x5}, or Obukhov \cite{x11}. We denote this class of models as \PG+. Sezgin and Niuwenhuizen \cite{x12} analyzed the particle spectrum of \PG+ in the weak field approximation around the Minkowski background $M_4$. Using the absence of ghosts and tachyons as physical requirements, they found a number of restrictions on the \PG+ parameters that ensure the propagating torsion modes to be well behaved.

General dynamical aspects of \PG+, including the identification of its
physical degrees of freedom, are most naturally understood in Dirac's
Hamiltonian approach for constrained dynamical systems \cite{x13}.
Blagojevi\'c and Nikoli\'c \cite{x14,x15} started a systematic Hamiltonian analysis of \PG+, focusing on its generic aspects. They identified a subset of the primary constraints that are always present (``sure" constraints,
associated to the local Poincar\'e symmetry). Moreover, if certain
critical parameters vanished, they found additional primary constraints (``if-constraints"), constructed the total Hamiltonian, and discussed certain aspects of the consistency procedure. Further advances in this direction were made by Cheng et al. \cite{x16} and Chen et al \cite{x17}, who found that the nonlinear nature of constraints may drastically change the number of propagating modes obtained in the linearized analysis. Yo and Nester \cite{x18} made a detailed study of this phenomenon in \PG+, concluding that there are apparently only two good propagating torsion modes. For an interesting application of this result to cosmology, see Shie et al. \cite{x19}.

There are no physical arguments that favor the conservation of parity in the
gravitational interaction. \emph{Parity violating models} based on the
general PG, with all possible quadratic invariants in the Lagrangian, were
considered already in the 1980s \cite{x20}, but the subject remained
without wider response. Recently, there has been increased interest in a
better understanding of both the basic structure and various dynamical
aspects of these  models, including cosmological applications and wave solutions \cite{x21,x22,x23,x24,x25,x26,x27}. In particular, one should
mention the analysis of the particle spectrum carried out by Karananas
\cite{x25}, who made a suitable extension of the weak field approximation
method used earlier in \PG+ \cite{x12} and applied it to the general PG. According to his results, it seems that the set of good modes that can coexist is significantly enlarged in comparison to \PG+.

The objective of the present work is to examine the Hamiltonian structure of the general PG, based on the if-constraint formalism \cite{x14,x15,x18}, and use it to clarify the physical content of its particle spectrum, calculated in the weak field approximation around $M_4$. In this regard, a particularly important role is played by both the critical parameters appearing in the analysis of the primary constraints, and the structure of the canonical Hamiltonian. By comparing the properties of the particle spectrum to those found in Ref. \cite{x25}, we noted certain differences. On the other side, elements of the Hamiltonian structure developed here can be a good starting point for studying the nonlinear dynamics of PG.

The paper is organized as follows. In Sec. \ref{sec2}, we give a short
account of the Lagrangian formalism for the general PG. In Secs.
\ref{sec3} and \ref{sec4}, we find the canonical critical parameters,
identify the related if-constraints, and construct the generic, ``most dynamical" canonical Hamiltonian, determined by the nonvanishing critical parameters. Then, in Secs. \ref{sec5} and \ref{sec6}, we derive the linearized gravitational field equations and use them to identify the mass eigenvalues of the torsion modes. The conditions for the absence of ghosts and tachyons, as well as the reality conditions of the mass eigenvalues, are examined in Sec. \ref{sec7}. Essential features of the particle spectrum are either tested by, or derived from the Hamiltonian structure of PG. In contrast to the results obtained in \cite{x25}, we show that the two spin-2 torsion modes cannot propagate simultaneously. In Sec. \ref{sec8}, we give a short summary of our results, and six appendices contain useful technical details, including an outline of the Hamiltonian formalism describing the case of vanishing critical parameters.

Our conventions are as follows. The Latin indices $(i,j,\dots)$ are the
local Lorentz indices, the Greek indices $(\m,\n,\dots)$ are the
coordinate indices, and both run over $0,1,2,3$; the orthonormal frame (tetrad) is $b^i{_\m}$, the inverse tetrad is $h_i{^\m}$, the Lorentz connection is $\om^{ij}{}_\m$,  $\eta_{ij}=(1,-1,-1,-1)$ and $g_{\m\n}=\eta_{ij} b^i{_\m}b^j{_\n}$ are the metric components in the local Lorentz and coordinate frame, respectively; a totally antisymmetric tensor $\ve_{ijkl}$ is normalized to $\ve_{0123}=+1$, and the dual of an antisymmetric tensor $X_{ij}$ is $\hd X_{ij}= (1/2)\ve_{ij}{}^{mn}X_{mn}$.

\section{Lagrangian formalism}\label{sec2}
\setcounter{equation}{0}

In this section, we give a short account of the Lagrangian formalism
for the general parity violating PG. Basic dynamical variables are the
tetrad field $b^i=b^i{_\m}dx^\m$ and the antisymmetric spin connection
$\om^{ij}=\om^{ij}{_\m}dx^\m=-\om^{ji}$ (1-forms), which represent the
gauge potentials associated with translations and Lorentz transformations,
respectively. The corresponding field strengths are the torsion and the
curvature (2-forms),
\bea
&&T^i:=d b^i+\om^i{_k}\wedge b^k
      =\frac{1}{2}T^i{}_{\m\n}dx^\m\wedge dx^\n \, ,             \nn\\
&&R^{ij}:=d\om^{ij}+\om^i{}_k\wedge\om^{kj}
         =\frac{1}{2}R^{ij}{}_{\m\n}dx^\m\wedge dx^\n\, ,
\eea
which satisfy the Bianchi identities
\be
\nab T^i=R^i{}_k\wedge b^k\,,\qquad \nab R^{ij}=0\, .                 \lab{2.2}
\ee
The underlying spacetime continuum is described by Riemann-Cartan
geometry \cite{x7,x8,x9}.

\subsection{Field equations}

The PG dynamics is determined by a Lagrangian $L=L_M+L_G$, where $L_M$
describes matter and its interaction with gravity, and $L_G$ is the pure
gravitational part. In the framework of tensor calculus, the gravitational field equations in vacuum are obtained by varying the action
$I_G=\int d^4 x L_G(b^i{_\m},T_{ijk},R_{ijkl})$ with respect to $b^i{_\m}$ and
$\om^{ij}{_\m}$. After introducing the covariant gravitational momenta
\bsubeq
\be
H_i{}^{\m\n}:=\frac{\pd L_G}{\pd T^i{}_{\m\n}}\, ,\qquad
H_{ij}{}^{\m\n}:=\frac{\pd L_G}{\pd R^{ij}{}_{\m\n}}\,,         \lab{2.3a}
\ee
and the associated energy-momentum and spin currents
\be
E_i{^\n}:=\frac{\pd L_G}{\pd b^i{_\m}}\,,\qquad
  E_{ij}{}^\m:=\frac{\pd L_G}{\pd\om^{ij}{_\m}}\,,              \lab{2.3b}
\ee
\esubeq
the gravitational field equations take a compact form
\bsubeq\lab{2.4}
\bea
\text{(1ST)}
&&\cE_i{^\n}:=-\frac{\d L_G}{\d b^i{_\m}}
             =\nab_\m H_i{}^{\m\n}-E_i{^\n}=0\, ,               \lab{2.4a}\\
\text{(2ND)}
&&\cE_{ij}{^\n}:=-\frac{\d L_G}{\d\om^{ij}{_\m}}
                =\nab_\m H_{ij}{}^{\m\n}-E_{ij}{^\n}=0\, .      \lab{2.4b}
\eea
\esubeq
The explicit expressions for the energy-momentum and spin currents are given by
\bea
&&E_i{^\n}=h_i{^\n}L_G
   -T^m{}_{ki}H_m{}^{k\n}-\frac{1}{2}R^{mn}{}_{ki}H_{mn}{}^{k\n}\,,\nn\\
&&E_{ij}{}^\m=-2H_{[ij]}{^\m}\,.
\eea

In the presence of matter, the right-hand sides of \eq{2.4a} and \eq{2.4b} contain the corresponding matter currents.

\subsection{Quadratic PG models}

We assume the Lagrangian density $L_G$ to contain all possible quadratic
invariants, constructed out of the three irreducible components of the torsion,
and the six irreducible components of the curvature (Appendix \ref{appA}). Relying on the Lagrangian 4-form given in Ref. \cite{x27}, one finds that the corresponding Lagrangian density has the form $L_G=b\cL_G$, where $b:=\det (b^i{_\m})$ and
\bea
\cL_G&=&-(a_0R+2\L_0)-\bara_0 X                                    \nn\\
&&+\frac{1}{2}T^{ijk}\sum_{n=1}^3
   \left(a_n\irr{n}T_{ijk}-\bara_n\hd\irr{n}{T}_{ijk}\right)\, ,   \nn\\
&&+\frac{1}{4}R^{ijkl}\sum_{n=1}^6
   \left(b_n\irr{n}{R}_{ijkl}-\barb_n\hd\irr{n}{R}_{ijkl}\right)\,.\lab{2.6}
\eea
Here, the irreducible components of the field strengths are defined in Appendix \ref{appA}, the parity even and parity odd sectors are described by the parameters $(a_n,b_n,\L_0)$ and $(\bara_n,\barb_n)$, respectively, and the star
symbol denotes the duality operation with respect to the frame indices of the field strengths. Another form of $\cL_G$, useful for comparison with the literature, is given in Appendix \ref{appB}. Knowing $\cL_G$, one finds that the covariant momentum densities \eq{2.3a} can be written in the form $H_{imn}=b\cH_{imn}$ and $H_{ijmn}=b\cH_{ijmn}$, where
\bsubeq
\bea\lab{2.7}
&&\cH_{imn}=2\sum_{n=1}^3\left(a_n\irr{n}T_{imn}
                               -\bara_n\hd T_{imn}\right)\,,    \nn\\
&&\cH_{ijmn}=\ir{L}\,\cH_{ijmn}+\cH'_{ijmn}\, ,
\eea
and
\bea
&&\ir{L}\,\cH_{ijmn}=-2a_0(\eta_{im}\eta_{jn}-\eta_{jm}\eta_{in})
                    +2\bara_0\ve_{ijmn}\, ,                     \nn\\
&&\cH'_{ijmn}=2\sum_{n=1}^6\left(b_n\irr{n}{R}_{ijmn}
              -\barb_n\hd\irr{n}{R}_{ijmn}\right)\, .
\eea
\esubeq

\subsection{On the choice of Lagrangian parameters}

In the Lagrangian \eq{2.6}, the two parity sectors are presented in a very
symmetric way, but the set of three identities \eq{A.3a} implies that not all of the parameters $(\bara_n,\barb_n)$ are independent. To resolve this issue, we choose the conditions
\be
\bara_2=\bara_3\, ,\qquad \barb_2=\barb_4\,,\qquad \barb_3=\barb_6\,,\lab{2.8}
\ee
which reduce the number of Lagrangian parameters to $21-3=18$. Note that
the above conditions are not unique.

Further freedom in the choice of parameters follows from the existence of
three topological invariants. The Euler and Pontryagin invariants are defined
by the 4-forms
\bsubeq\lab{2.9}
\be
I_E:=R^{ij}\wedge R^{mn}\ve_{mnij}\, ,\qquad I_P:=R^{ij}\wedge R_{ij}\,,
\ee
respectively, whereas the third invariant is based on the Nieh-Yan
identity,
\be
I_{NY}:=T^i\wedge T_i-R_{ij}\wedge b^i\wedge b^j\equiv d(b^i\wedge T_i)\, .
\ee
\esubeq
These invariants produce vanishing contributions to the field equations,
which implies that not all of the Lagrangian parameters are dynamically
independent. Indeed, they can be used to eliminate three more terms from the
Lagrangian, leaving us with the final number of $18-3=15$ independent
parameters; see Ref. \cite{x23}$_3$ for more details. In this paper, we
use only the conditions \eq{2.8}, allowing thereby for an easier comparison to the literature.

For a clear understanding of the physical content of PG, it is convenient  to use dimensionless parameters (coupling constants). The Lagrangian parameters in \eq{2.6} are not dimensionless, but the transition to their dimensionless counterparts can be easily realized by suitable rescalings; see for instance Ref. \cite{x27}. However, to make the general exposition simpler and more compact, we find it useful to keep the Lagrangian parameters in the form \eq{2.6}, which corresponds to using the units $c=\hbar=2\k=1$. The true dimensionless parameters can be reintroduced later whenever needed.

\section{Primary constraints}\label{sec3}
\setcounter{equation}{0}

Hamiltonian structure is by itself a particularly important aspect of PG as a gauge theory \cite{x13}. Moreover, it also offers dynamical information that is essential for a proper understanding of the particle spectrum of PG.

We begin the subject by analyzing the primary constraints (PC) of PG. The canonical momenta $(\pi_i{^\m},\Pi_{ij}{^\m})$ associated to the basic Lagrangian variables $(b^i{_\m},\om^{ij}{_\m})$ are
\be\lab{3.1}
\pi_i{^\m}:=\frac{\pd L_G}{\pd(\pd_0 b^i{_\m})}=b\cH_i{}^{0\m}\, ,\qquad
\Pi_{ij}{^\m}:=\frac{\pd L_G}{\pd(\pd_0\om^{ij}{_\m})}=b\cH_{ij}{}^{0\m}\, .
\ee
Since the field strengths do not depend on the velocities $\pd_0 b^i{_0}$
and $\pd_0\om^{ij}{_0}$, the above relations define 10 constraints that are always present in the theory (``sure" PCs), regardless of the values of the coupling constants. They read
\be
\pi_i{^0}\approx 0\,,\qquad \Pi_{ij}{^0}\approx 0\, .           \lab{3.2}
\ee
and their existence is directly related to the local Poincar\'e symmetry of PG.

Before we proceed, let us note that at each point of a spatial hypersurface $\S_0: x^0=$const., one can define the unit timelike vector \mb{n}, normal to $\S_0$. Then, any spacetime vector $V_k$ can be decomposed into a component $V_\orth$ along \mb{n}, and a component $V_\bk$ in the tangent space of (``parallel" to) $\S_0$; that is $V_k=n_k V_\orth +V_\bk$, where $V_\orth=n^k V_k$ and $n^kV_\bk=0$ (Appendix \ref{appC}).

To find additional constraints that may appear in \eq{3.1}, it is useful
to define the parallel gravitational momenta
\bsubeq\lab{3.3}
\bea
&&\hpi_{i\bk}:=\pi_i{^\a}b_{k\a}=J\cH_{i\orth\bk}\, ,            \lab{3.3a}\\
&&\hPi_{ij\bk}:=\Pi_{ij}{^\a}b_{k\a}=J\cH_{ij\orth\bk}\, ,       \lab{3.3b}
\eea
\esubeq
such that $\hpi_{i\bk}n^k=0$, $\Pi_{ij\bk}n^k=0$, and $J$ is defined by
$b=NJ$, with $N=n_k b^k{_0}$. Depending on the values of the coupling
constants, these relations may produce additional constraints (primary
``if-constraints"). In analogy to the above orthogonal-parallel
decomposition of a vector $V_k$, one can introduce a similar decomposition
of the field strengths,
\bsubeq\lab{3.4}
\bea
&&T_{ikm}=T_{i\bk\bm}+(n_k T_{i\orth\bm}+n_mT_{i\bk\orth})
         =\bT_{ikm}+\cT_{ikm}\, ,                               \lab{3.4a}\\
&&R_{ijkm}=R_{ij\bk\bm}+(n_kR_{ij\orth\bm}+n_m R_{ij\bk\orth})
          =\bR_{ijkm}+\cR_{ijkm}\, .                            \lab{3.4b}
\eea
\esubeq
It is very useful for further analysis to know that the parallel
components $\bT_{ikm}:=T_{i\bk\bm}$ and $\bR_{ijkm}:=R_{ij\bk\bm}$ are
independent not only of the ``velocities" $T_{i\orth\bm},R_{ij\orth\bm}$,
but also of the unphysical variables $(b^i{_0},\om^{ij}{_0})$; for more
details, see Refs. \cite{x7,x14,x15}.

\subsection{Torsion sector}

The torsion piece of the Lagrangian \eq{2.6} depends on the velocities $\pd_0 b^i{_\a}$ only through $T_{i\orth\bk}$. The linearity of $\cH_{i\orth\bk}$ in $\bT$ and $\cT$ allows us to rewrite \eq{3.3a} in the form
\be
\phi_{i\bk}:=\frac{\hpi_{i\bk}}{J}-\cH_{i\orth\bk}(\bT)
            =\cH_{i\orth\bk}(\cT)\, ,                           \lab{3.5}
\ee
where all possible velocity terms are moved to the right-hand side.
Now, we decompose this equation into irreducible parts with respect to the group of rotations in $\S_0$ (Appendix \ref{appC}):
\bsubeq\lab{3.6}
\bea
\ir{S}\phi&:=&\frac{\hpi_\bk{^\bk}}{J}+\bara_2\ve^{\bk\bm\bm}T_{\bk\bm\bn}
               =-2a_2T^\bk{}_{\bk\orth}\, ,                     \lab{3.6a}\\
\phi_{\orth\bk} &:=&\frac{\hpi_{\orth\bk}}{J}
  +\frac{2}{3}(a_1-a_2)T^\bm{}_{\bm\bk}
  +\frac{1}{3}(2\bara_1+\bara_2)\ve_\bk{}^{\bm\bn}T_{\orth\bm\bn}\nn\\
  &=&\frac{2}{3}(2a_1+a_2)T_{\orth\orth\bk}
     +\frac{2}{3}(\bara_1-\bara_2)
                      \ve_\bk{}^{\bm\bn}T_{\bm\bn\orth}\,,      \lab{3.6b}\\
\ir{A}\phi_{\bi\bk}&:=&\frac{^A\hpi_{\bi\bk}}{J}
     -\frac{2}{3}(a_1-a_3)T_{\orth\bi\bk}
     -\frac{1}{3}(\bara_1+2\bara_3)\ve_{\bi\bk}{}^\bn T^\bm{}_{\bm\bn}\nn\\
  &=&-\frac{2}{3}(a_1+2a_3)T_{[\bi\bk]\orth}
     -\frac{2}{3}(\bara_1-\bara_2)
                   \ve_{\bi\bk}{}^\bn T_{\orth\orth\bn}\,,      \lab{3.6c}\\
\ir{T}\phi_{\bi\bk}&:=&\frac{^T\hpi_{\bi\bk}}{J}
   +\bara_1\left[\ve_{(\bi}{}^{\bm\bn}T_{\bk)\bm\bn}
           -\frac{1}{3}\eta_{\bi\bk}\ve^{\bk\bm\bn}T_{\bk\bm\bn}\right]
                         =-2a_1{}^TT_{\bi\bk\orth}\, .          \lab{3.6d}
\eea
\esubeq
Here, the set $(\ir{S}\phi,\phi_{\orth\bk},\ir{A}\phi_{\bi\bk},\ir{T}\phi_{\bi\bk})$,
defined by the scalar, vector, antisymmetric and traceless-symmetric parts
of $\phi_{i\bk}$, represents the set of all possible new constraints. The
mechanism by which these if-constraints become true constraints is simply
explained in the parity even case, characterized by four critical parameters:
$a_2,(2a_1+a_2),(a_1+2a_3)$, and $a_1$. When some of these parameters
vanish, the corresponding velocity terms on the right-hand sides of \eq{3.6} also vanish, and consequently, the associated if-constraints become true constraints. However, if none of the critical parameters vanishes, there are no new constraints.

The same mechanism works also in the general PG. Whereas the critical
parameters for $\ir{S}\phi$ and $\ir{T}\phi_{\bi\bk}$ remain the same as
in \PG+, $a_2$ and $a_1$, the structure of the if-constraints
$\phi_{\orth\bk}$ and $\ir{A}\phi_{\bi\bk}$ is more complicated, as the
right-hand sides of \eq{3.6b} and \eq{3.6c} depend on two velocities, $T_{[\bm\bn]\orth}$ and $T_{\orth\orth\bk}$. To find the related critical parameters, we first transform $\ir{A}\phi_{\bi\bk}$ into the axial 3-vector
$\ir{A}\phi_\bk:=\ve_\bk{}^{\bm\bn}\ir{A}\phi_{\bm\bn}$, so that \eq{3.6c} goes over into
\be
\ir{A}\phi_\bk =\frac{4}{3}(\bara_1-\bara_2)T_{\orth\orth\bk}
     -\frac{2}{3}(a_1+2a_3)\ve_\bk{}^{\bm\bn}T_{\bm\bn\orth}\,. \lab{3.7}
\ee
Then, the set of equations involving
$(\phi_{\orth\bk},\ir{A}\phi_\bk)$ can be written in the matrix form as
\bsubeq\lab{3.8}
\be
\left(\ba{c}
      \phi_{\orth\bk} \\
      \ir{A}\phi_\bk
      \ea
      \right)=\frac{2}{3}A \left(\ba{c}
                      T_{\orth\orth\bk} \\
                      \ve_\bk{}^{\bm\bn}T_{\bm\bn\orth}
                      \ea
                      \right)\, ,                               \lab{3.8a}
\ee
where
\bea
&&A:=\left( \ba{cc}
          2a_1+a_2           & \bara_1-\bara_2 \\
          2(\bara_1-\bara_2) & -(a_1+2a_3)
          \ea
         \right)\, ,                                            \nn\\
&&\det A=-\Big[(2a_1+a_2)(a_1+2a_3)+2(\bara_1-\bara_2)^2\Big]\,.\lab{3.8b}
\eea
\esubeq
If the matrix $A$ has two distinct eigenvalues, one can construct the invertible matrix $P$ that transforms $A$ into a diagonal form, $D_A:=P^{-1}AP$. Then, Eq. \eq{3.8a} implies
\be
\phi_\bk  :=  P^{-1}\left(\ba{c}
                      \phi_{\orth\bk} \\
                      \ir{A}\phi_\bk
                          \ea\right)
                =\frac{2}{3}D_AP^{-1}\left(\ba{c}
                            T_{\orth\orth\bk} \\
                            \ve_\bk{}^{\bm\bn}T_{\bm\bn\orth}
                                       \ea\right)\,,          \lab{3.9}
\ee
where the column $\phi_\bk$ represents two diagonalized if-constraints, and the
diagonal elements of $D_A$ are the critical parameters,
\be
c_\pm(A)=\frac{1}{2}\Big(\tr A\pm\sqrt{(\tr A)^2-4\det A}\,\Big)\, .
\ee
More details on this construction can be found in Appendix \ref{appD}. In general, the number of true constraints in \eq{3.9} is equal to the number of vanishing critical parameters.
\bitem
\item[$\bull$] The critical parameters of the torsion sector are $a_2,c_\pm(A)$, and $a_1$.
\eitem

\subsection{Curvature sector}

For the curvature sector, we use the linearity of $\cH'_{ij\orth
k}$ in $\bR$ and $\cR$ to rewrite \eq{3.3b} in the form
\be
\Phi_{ij\bk}:=\frac{\Pi_{ij\bk}}{J}-\cH_{ij\orth\bk}(\bR)
             =\cH'_{ij\orth\bk}(\cR)\, .
\ee
The content of the object $\Phi_{ij\bk}$ is described by two three-dimensional (3d) tensors, $\Phi_{ij\bk}=(\Phi_{\orth\bj\bk},\Phi_{\bi\bj\bk})$. The irreducible decomposition of $\Phi_{\orth\bj\bk}$ takes the form defined in \eq{C.4}:
\bsubeq\lab{3.12}
\bea
\ir{S}\Phi&\equiv&\frac{\Pi_{\orth\bk}{}^\bk}{J}+6a_0
   +\frac{1}{2}(b_4-b_6)\ul{R}
   +\frac{1}{2}(\barb_2+\barb_3)\ve^{\bk\bm\bn}R_{\orth\bk\bm\bn} \nn\\
  &=&(b_4+b_6)R_{\orth\orth}
 -\frac{1}{2}(\barb_2-\barb_3)\ve^{\bk\bm\bn}R_{\bk\bm\bn\orth}\,,\lab{3.12a}\\
\ir{A}\Phi_{\orth\bj\bk}&\equiv& \frac{\ir{A}\Pi_{\orth\bj\bk}}{J}
  +(b_2-b_5)^A\,\ul{R}_{\bj\bk}
  -\frac{1}{2}(\barb_2+\barb_5)\,
          \ir{A}\Big(\ve_\bj{}^{\bm\bn}R_{\orth\bk\bm\bn}\Big)    \nn\\
  &=&(b_2+b_5)\ir{A}R_{\orth\bj\orth\bk}
     -\frac{1}{2}(\barb_2-\barb_5)\,
      \ir{A}\Big(\ve_\bj{}^{\bm\bn}R_{\bm\bn\bk\orth}\Big)\,,     \lab{3.12b}\\
\ir{T}\Phi_{\orth\bj\bk}&\equiv& \frac{\ir{T}\Pi_{\orth\bj\bk}}{J}
  +(b_1-b_4)\,\ir{T}\ul{R}_{\bj\bk}
  +\frac{1}{2}(\barb_1+\barb_2)\,
          \ir{T}\Big(\ve_\bj{}^{\bm\bn}R_{\orth\bk\bm\bn}\Big)    \nn\\
  &=&(b_1+b_4)\,\ir{T}R_{\orth\bj\orth\bk}
      -\frac{1}{2}(\barb_1-\barb_2)
      \ir{T}\Big(\ve_\bj{}^{\bm\bn}R_{\bm\bn\orth\bk}\Big)\,.    \lab{3.12c}
\eea
\esubeq
The irreducible parts of $\Phi_{\bi\bj\bk}=-\Phi_{\bj\bi\bk}$ are the
pseudoscalar, the vector and the traceless symmetric part, see \eq{C.5}:
\bsubeq\lab{3.13}
\bea
\ir{P}\Phi&\equiv&
     \frac{\ir{P}\Pi}{J}+12\bara_0
             +(b_2-b_3)\ve^{\bk\bm\bn}R_{\orth\bk\bm\bn}
                           -(\barb_1+2\barb_2+\barb_3)\ul{R}    \nn\\
 &=&-(b_2+b_3)\ve^{\bk\bm\bn}R_{\bk\bm\bn\orth}
         -2(\barb_2-\barb_3)R_{\orth\orth}\, ,                  \lab{3.13a}\\
\ir{V}\Phi_\bi&\equiv&
  \frac{\ir{V}\Pi_\bi}{J}-(b_4-b_5)R_{\orth\bi}
  +\frac{1}{2}(\barb_2+\barb_5)\ve^{\bk\bm\bn}R_{\bi\bk\bm\bn}  \nn\\
 &=&(b_4+b_5)R_{\bi\orth}-(\barb_2-\barb_5)
       \ve_\bi{}^{\bk\bn}R_{\orth\bk\orth\bn}\,,                \lab{3.13b}\\
\ir{T}\Phi_{\bi\bj\bk}&\equiv&\frac{\ir{T}\Pi_{\bi\bj\bk}}{J}
   +(b_1-b_2)\ir{T}R_{\orth\bj\bk\bi}
                   -\ir{T}\cH^{\prime-}_{\bi\bj\orth\bk}(\bR)   \nn\\
 &=&(b_1+b_2)\ir{T}R_{\bi\bj\orth\bk}
   -(\barb_1-\barb_2)\ir{T}\Bigl(
   \ve_{\bi\bj}{^\bn}\,\ir{\S}R_{\orth\bn\orth\bk}\Big)\,.      \lab{3.13c}
\eea
\esubeq
In Eqs. \eq{3.12} and \eq{3.13}, the underlined objects do not contain velocities, $\ul{R}:=R^{\bm\bn}{}_{\bm\bn}$ and
$\ul{R}_{\bi\bj}:=R_{\bi\bn}{}_\bj{^\bn}$, the superscript $\S$ denotes symmetrization, and $\cH^{\prime-}_{\bi\bj\orth\bk}$ in \eq{3.13c} denotes the parity odd part of the covariant momentum,
$$
\ir{T}\cH^{\prime-}_{\bi\bj\orth\bk}(\bR)=-\ir{T}\left\{\frac{1}{4}
        \ve_\bk{}^{\bm\bn}\Big[(\barb_1+2\barb_2+\barb_5)R_{\bi\bj\bm\bn}
                    +(\barb_1-\barb_5)R_{\bm\bn\bi\bj}\Big]\right\}\, .
$$

Looking at the type of velocities appearing in the above equations, one
can see that the critical parameters can be found by grouping these
equations into suitably chosen pairs.

\subsubsection*{Spin-0 pair}

Consider first Eqs. \eq{3.12a} and \eq{3.13a}, which contain the same set of
velocities, $R_{\orth\orth}$ and $\ve^{\bk\bm\bn}R_{\bk\bm\bn\orth}$. They
can be written in the matrix form as
\bsubeq
\be
\left(\ba{c}
      \ir{S}\Phi \\
      \ir{P}\Phi
      \ea
      \right)=B_0\left(\ba{c}
                      R_{\orth\orth} \\
                      \ve^{\bk\bm\bn}R_{\bk\bm\bn\orth}
                      \ea
                      \right)\, ,                               \lab{3.14a}
\ee
where
\bea
B_0&:=&\left( \ba{cc}
          b_4+b_6    & -\frac{1}{2}(\barb_2-\barb_3)\\
          -2(\barb_2-\barb_3) & -(b_2+b_3)
          \ea
         \right)\, ,                                            \nn\\
\det B_0&=&-(b_4+b_6)(b_2+b_3)-(\barb_2-\barb_3)^2\, .          \lab{3.14b}
\eea
\esubeq
In analogy to what we found in the previous subsection, the critical
parameters are the eigenvalues of $B_0$, $c_\pm(B_0)$, and the related column of the if-constraints reads
\be
\ir{0}\Phi   := P_0^{-1}\left(\ba{c}
                            \ir{S}\Phi \\
                            \ir{P}\Phi
                              \ea\right)
  =D_0P_0^{-1}\left(\ba{c}
                      R_{\orth\orth} \\
                      \ve^{\bk\bm\bn}R_{\bk\bm\bn\orth}
                      \ea
                      \right)\, ,                               \lab{3.15}
\ee
were $P_0$ is the matrix that diagonalizes $B_0$, $D_0=P_0^{-1}B_0P_0$.

\subsubsection*{Spin-1 pair}

Similarly, after transforming  $\ir{A}\Phi_{\orth\bi\bj}$ into
$\ir{A}\Phi_\bk:=\ve_\bk{}^{\bm\bn}\Phi_{\orth\bm\bn}$, Eq. \eq{3.12b} becomes
\be
\ir{A}\Phi_\bi=(b_2+b_5)\ve_\bi{}^{\bm\bn}R_{\orth\bm\orth\bn}
                +(\barb_2-\barb_5) R_{\bi\orth}\, ,            \lab{3.16}
\ee
and the matrix form of Eqs. \eq{3.16} and \eq{3.13b} reads
\bsubeq
\be
\left(\ba{c}
      \ir{A}\Phi_\bi \\
      \ir{V}\Phi_\bi
      \ea
      \right)=B_1 \left(\ba{c}
                      \ve_\bi{}^{\bm\bn}R_{\orth\bm\orth\bn} \\
                      R_{\bi\orth}
                      \ea
                      \right)\, ,                               \lab{3.17a}
\ee
where
\bea
B_1&:=&\left( \ba{cc}
          b_2+b_5    & \barb_2-\barb_5 \\
          -(\barb_2-\barb_5) & b_4+b_5
          \ea
         \right)\, ,                                            \nn\\
\det B_1&=&(b_4+b_5)(b_2+b_5)+(\barb_2-\barb_5)^2\, .           \lab{3.17b}
\eea
\esubeq
As before, the critical parameters are $c_\pm(B_1)$, and the
if-constraints are determined by the matrix $P_1$ that diagonalizes $B_1$,
\be
\ir{1}\Phi_\bi
   := P_1^{-1}\left(\ba{c}
           \ir{A}\Phi_\bi \\
           \ir{V}\Phi_\bi
               \ea \right)
    =D_1P_1^{-1}\left(\ba{c}
                      \ve_\bi{}^{\bm\bn}R_{\orth\bm\orth\bn} \\
                      R_{\bi\orth}
                      \ea
                      \right)\, .                               \lab{3.18}
\ee

\subsubsection*{Spin-2 pair}

To find the critical parameters in the spin-2 sector, it is convenient to replace $\ir{T}\Phi_{\bi\bj\bk}$ by the expression $\ir{T}\Phi_{\bi\bk}:=\ir{T}\big(\ve_\bi{}^{\bm\bn}\Phi_{\bm\bn\bk}\big)$. Indeed, $\ir{T}\Phi_{\bi\bk}$ refers to the same set of velocities that appears in Eq. \eq{3.12c},
\be
\ir{T}\Phi_{\bi\bk}
      =(b_1+b_2)\ir{T}\big(\ve_\bi{}^{\bm\bn}R_{\bm\bn\orth\bk}\big)
        +2(\barb_1-\barb_2)\ir{T}R_{\orth\bi\orth\bk} \, ,      \lab{3.19}
\ee
which allows Eqs. \eq{3.12c} and \eq{3.19} to be written in the matrix form
\bsubeq
\be
\left(\ba{c}
      \ir{T}\Phi_{\orth\bj\bk} \\
      \ir{T}\Phi_{\bj\bk}
      \ea
      \right)=B_2\left(\ba{c}
                 \ir{T}R_{\orth\bj\orth\bk}\\
                 \ir{T}\big(\ve_\bj{}^{\bm\bn}R_{\bm\bn\orth\bk}\big)
                       \ea
                       \right)\, ,                              \lab{3.20a}
\ee
where
\bea
B_2&:=&\left( \ba{cc}
          b_1+b_4    & -\frac{1}{2}(\barb_1-\barb_2) \\[3pt]
          2(\barb_1-\barb_2) & b_1+b_2
          \ea
         \right)\, ,                                            \nn\\
\det B_2&=& (b_1+b_2)(b_1+b_4)+(\barb_1-\barb_2)^2\, .          \lab{3.20b}
\eea
\esubeq
Hence, the critical parameters are $c_\pm(B_2)$, and the column of if-constraints has the form
\be
\ir{2}\Phi_{\bj\bk}:=P_2^{-1}\left(\ba{c}
                       \ir{T}\Phi_{\orth\bj\bk} \\
                       \ir{T}\Phi_{\bj\bk}
                              \ea
                             \right)
   = D_2P_2^{-1}\left(\ba{c}
                 \ir{T}R_{\orth\bj\orth\bk}\\
                 \ir{T}\big(\ve_\bj{}^{\bm\bn}R_{\bm\bn\orth\bk}\big)
                       \ea
                       \right)\, .                              \lab{3.21}
\ee

\bitem
\item[$\bull$] The critical parameters in the curvature sector are
$c_\pm(B_0),c_\pm(B_1)$, and $c_\pm(B_2)$.
\eitem

\subsection{Critical parameters and if-constraints}

Since the if-constraints belong to irreducible representations of 3d
rotations, they are characterized by a specific spin content. Their
structure is best understood by grouping them into pairs with definite
spin, as shown in Table 1. In this classification, the parity eigenvalues
are absent since parity is not conserved.

\begin{center}
\doublerulesep 1.8pt
\begin{tabular}{lll}
\multicolumn{3}{l}{\hspace{16pt}Table 1. Critical parameters and
                                                 if-constraints} \\
                                                          \hline\hline
\rule{0pt}{12pt}
Spin &~Critical parameters &~ If-constraints                 \\
                                                             \hline
\rule[-1pt]{0pt}{15pt}
\ph{x} 0  &~$a_2,\,c_\pm(B_0)$
                 &~$\ir{S}\phi,(\ir{0}\Phi)_\pm$               \\
\rule[-1pt]{0pt}{15pt}
\ph{x} 1  &~$c_\pm(A),\, c_\pm(B_1)$
              &~$(\phi_\bk)_\pm,(\ir{1}\Phi_\bk)_\pm$  \\[2pt]
\rule[-1pt]{0pt}{15pt}
\ph{x} 2  &~$a_1,\,c_\pm(B_2)$
        &~$\ir{T}\phi_{\bi\bk},(\ir{2}\Phi_{\bi\bk})_\pm$ \\[2pt]
                                                         \hline\hline
\end{tabular}
\end{center}
The generic set of the critical parameters $c_\pm(F)$, $F=A,B_0,B_1,B_2$, is defined provided the parity odd parameters in $F$ do not vanish, see Appendix \ref{appD}. Hence, the limit of the final expressions $c_\pm(F)$ when these parameters tend to zero is not well defined. However, since in that case $F$ is already diagonal, one can identify $c_\pm$ directly from $F$.

The total number of the primary if-constraints is $10\times 3=30$, the same as
the number of the parallel canonical momenta \eq{3.3}.
The if-constraints and the associated critical parameters have a decisive influence on the structure of the canonical Hamiltonian.

\section{Hamiltonian}\label{sec4}
\setcounter{equation}{0}

The procedure for constructing the canonical (and total) Hamiltonian in \PG+ is well known  \cite{x7,x14,x15,x18}, but its extension to PG, although in principle straightforward, is technically rather complicated.

Starting with the standard definition of the canonical Hamiltonian
density,
\be
\cH_c=\p_i{^\a}\pd_0 b^i{_\a}
      +\frac{1}{2}\Pi_{ij}{^\a}\pd_0\om^{ij}{_\a}-b\cL\, ,      \lab{4.1}
\ee
one can rewrite it in the Dirac-ADM form:
\be
\cH_c=N\cH_\orth+N^\a\cH_\a
       -\frac{1}{2}\om^{ij}{_0}\cH_{ij}+\pd_\a D^\a\, ,         \lab{4.2}
\ee
where $N$ and $N^\a$ are the lapse and shift functions (see Appendix \ref{appC}), and
\bea
&&\cH_{ij}=2\pi_{[i}{}^\a b_{j]\a}+\nab_\a\Pi_{ij}{^\a}\, ,     \nn\\
&&\cH_\a=\pi_i{^\b}T^i{}_{\a\b}
         +\frac{1}{2}\pi_{ij}{^\b}R^{ij}{}_{\a\b}
                          -b^k{_\a}\nab_\b\pi_k{^\b}\, ,        \nn\\
&&\cH_\orth=\pi_i{^\bk}T^i{}_{\orth\bk}
         +\frac{1}{2}\Pi_{ij}{}^\bk R^{ij}{}_{\orth\bk}
                          -J\cL-n^k\nab_\b\pi_k{^\b}\,,         \nn\\
&&D^\a=b^i{_0}\p_i{^\a}+\frac{1}{2}\om^{ij}{_0}\Pi_{ij}{^\a}\,. \lab{4.3}
\eea
Since $\cH_\orth$ is the only term that depends on the form of the
Lagrangian, explicit construction of the whole $\cH_c$ reduces just to the
construction of its dynamical piece $\cH_\orth$. In this process, we focus
our attention on the ``most dynamical" case when all the critical parameters are nonvanishing (that is, when none of the if-constraints becomes a true constraint). Such an assumption is sufficient for our study of the particle spectrum of PG. Extension of the formalism to include vanishing critical parameters is outlined in Appendix \ref{appD}.

\subsection{Torsion sector}\label{sub41}

Isolating the torsion contribution to $\cL_G$, one finds the
corresponding part of $\cH_\orth$,
\bsubeq
\be
\cH_\orth^T=\frac{1}{2}\phi_{i\orth\bk}T^{i\orth\bk}
              -J\bcL_{T^2}-n^k\nab_\b\pi_k{^\b}\, ,
\ee
where $\bcL_{T^2}=\cL_{T^2}(\bT)$ does not contain velocities. In order to
express the velocities in terms of the phase-space variables, we decompose
the first term into four irreducible parts:
\be
\phi_{i\orth\bk}T^{i\orth\bk}=\phi_{\orth\bk}T^{\orth\orth\bk}
  +\frac{1}{2}\ir{A}\phi_\bi\,\ve^{\bi}{}_{\bm\bn}T^{\bm\bn\orth}
  +\ir{T}\phi_{\bi\bk}\ir{T}T^{\bi\orth\bk}
  +\frac{1}{3}\ir{S}\phi T_{\bk\orth}{}^\bk\, .                 \lab{4.4b}
\ee
\esubeq
If $a_1,a_2\ne 0$, the velocities from the last two terms can be directly
eliminated using Eqs. \eq{3.6a} and \eq{3.6d},
\bsubeq
\be
\frac{1}{3}\ir{S}\phi T_{\bk\orth}{}^\bk
+\ir{T}\phi_{\bi\bk}\ir{T}T^{\bi\orth\bk}=
  \frac{1}{6a_2}\ir{S}\phi\ir{S}\phi
  +\frac{1}{2a_1}\ir{T}\phi_{\bi\bk}\ir{T}\phi^{\bi\bk}\,.      \lab{4.5a}
\ee
Continuing with the first two terms in \eq{4.4b}, we note that, for $\det A\ne 0$, one can use the relation $A^{-1}\times$\eq{3.8a} to eliminate the velocities. Introducing the notation $\vphi_\bk:=(\phi_{\orth\bk},\ir{A}\phi_\bk)^T$, the result takes a compact matrix form,
\bea
&&\left(\ba{ll}
      \phi_{\orth\bk}\,, & \frac{1}{2}\ir{A}\phi_\bk
        \ea\right) \left(\ba{c}
                    T^{\orth\orth\bk} \\
                    \ve^{\bk}{}_{\bm\bn}T^{\bm\bn\orth}
                    \ea\right)
   =\frac{3}{2\,\det A} \vphi^T_\bk T\vphi^\bk\, ,              \nn\\
&&T:=\left(\ba{cc}
           2a_1+a_2        & \bara_1-\bara_2 \\
           \bara_1-\bara_2 & -(a_1+2a_3)/2
           \ea\right)\, ,\qquad \det T=\frac{1}{2}\det A\, .
\eea
\esubeq
Hence, the resulting form of $\cH_\orth^T$ reads
\bea
&&\cH_\orth^T=\frac{1}{2}J\phi_T^2
                -J\bcL_{T^2}-n^k\nab_\b\pi_k{^\b}\, ,           \nn\\
&&\phi_T^2:=\frac{1}{6a_2}\ir{S}\phi\ir{S}\phi
  +\frac{1}{2a_1}\ir{T}\phi_{\bi\bk}\ir{T}\phi^{\bi\bk}
  +\frac{3}{2\,\det A}\vphi^T_\bk T\vphi^\bk\, ,                \lab{4.6}
\eea

\subsection{Curvature sector}

In a similar manner, one finds the curvature contribution to $\cH_\orth$:
\bsubeq
\be
\cH_\orth^R=\frac{1}{4}\Phi_{ij\bk}R^{ij\orth\bk}-J\bcL_{R^2}
   -a_0R^{\bm\bn}{}_{\bm\bn}+\bara_0\ve^{\bm\bn\bk}R_{\orth\bm\bn\bk}\, ,
\ee
where $\bcL_{R^2}=\cL_{R^2}(\bR)$ does not contain velocities, and
\bea
\Phi_{ij\bk}R^{ij\orth\bk}
&=&\frac{2}{3}\ir{S}\Phi R_{\orth\orth}
  -\ir{A}\Phi_\bk\ve^{\bk\bm\bn}R_{\orth\bm\orth\bn}
  +2\ir{T}\Phi_{\orth\bj\bk}\ir{T}R^{\orth\bj\orth\bk}\,,  \nn\\
&&-\frac{1}{6}\ir{P}\Phi\ve_{\bi\bj\bk}R^{\bi\bj\orth\bk}
  +\ir{V}\Phi^\bi R_{\bi\bk\orth}{^\bk}
  -\frac{1}{2}\ir{T}\Phi^{\bi\bk}
              \ir{T}(\ve_\bi{}^{\bm\bn}R_{\bm\bn\orth\bk})\, .  \lab{4.7b}
\eea
\esubeq

Summing up the scalar and pseudoscalar term from the expression \eq{4.7b}
and using the relation $B_0^{-1}\times$\eq{3.14a} to eliminate the
velocities, one obtains
\bea
&&\frac{2}{3}\ir{S}\Phi R_{\orth\orth}
   +\frac{1}{6}\ir{P}\Phi\ve_{\bi\bj\bk}R^{\bi\bj\bk\orth}
   =J\frac{1}{6\,\det B_0}\ir{(0)}\Phi^T R_0\ir{(0)}\Phi\,,   \lab{4.8}\\
&&R_0=\left(\ba{cc}
             -4(b_2+b_3)   & 2(\barb_2-\barb_3)  \\
        2(\barb_2-\barb_3) & b_4+b_6
            \ea\right)\, ,\qquad \det R_0=4\det B_0\,,        \nn
\eea
where $\ir{(0)}\Phi^T:=(\ir{S}\Phi,\ir{P}\Phi)$.

Similarly, the sum of the axial vector and vector term, combined with
$B_1^{-1}\times$\eq{3.17a}, yields
\bea
&&-\ir{A}\Phi_\bk\ve^{\bk\bm\bn}R_{\orth\bm\orth\bn}
  +\ir{V}\Phi^\bi R_{\bi\orth}=
  -J\frac{1}{\det B_1}\ir{(1)}\Phi_\bi^T R_1\ir{(1)}\Phi^\bi\,, \lab{4.9}\\
&&R_1=\left(\ba{cc}
             b_4+b_5 & -(\barb_2-\barb_5)  \\
        -(\barb_2-\barb_5) & -(b_2+b_5)
            \ea\right)\, ,\qquad \det R_1=-\det B_1\,,          \nn
\eea
where $\ir{(1)}\Phi^T_\bi=(\ir{A}\Phi_\bi,\ir{V}\Phi_\bi)$.

Finally, using $B_2^{-1}\times$\eq{3.20a}, the sum of the two tensor terms
is given by
\bea
&&2\ir{T}\Phi_{\orth\bj\bk}\ir{T}R^{\orth\bj\orth\bk}
   -\frac{1}{2}\ir{T}\Phi^{\bi\bk}\ir{T}(\ve_\bi{}^{\bm\bn}R_{\bm\bn\orth\bk})
   =J\frac{1}{4\,\det B_2} \ir{(2)}\Phi_{\bi\bk}^T
                           R_2\ir{(2)}\Phi^{\bi\bk}\,,         \lab{4.10}\\
&&R_2=\left(\ba{cc}
             4(b_1+b_2)    & 2(\barb_2-\barb_1)  \\
        2(\barb_2-\barb_1) & -(b_1+b_4)
            \ea\right)\, ,\qquad \det R_2=-4\det B_2\, ,       \nn
\eea
where~$\ir{(2)}\Phi^T_{\bi\bk}:=(\ir{T}\Phi_{\orth\bi\bk},\ir{T}\Phi_{\bi\bk})$.

Summing up the above three contributions, one obtains the expression for $\cH^R_\orth$ as
\bea
&&\cH^R_\orth=\frac{1}{4}J\Phi_R^2-J\cL_{R^2}(\ol{R})
    -a_0R^{\bm\bn}{}_{\bm\bn}
    +\bara_0\ve^{\bm\bn\bk}R_{\orth\bm\bn\bk}\, ,               \lab{4.11}\\
&&\Phi_R^2:=\frac{1}{6\,\det B_0}\ir{(0)}\Phi^T R_0\ir{(0)}\Phi
   -\frac{1}{\det B_1}\ir{(1)}\Phi_\bi^T R_1\ir{(1)}\Phi^\bi
   +\frac{1}{4\,\det B_2} \ir{(2)}\Phi_{\bi\bk}^T
                           R_2\ir{(2)}\Phi^{\bi\bk}\, .         \nn
\eea

The complete expression $\cH_\orth=\cH^T_\orth+\cH^R_\orth$ will be used in Sec. \ref{sec7} to formulate the conditions for the positivity of energy of the isolated spin modes.

\subsection{Consistency conditions}

The complete canonical Hamiltonian of PG, with
$\cH_\orth=\cH_\orth^T+\cH_\orth^R$, is calculated by assuming that
none of the critical parameters is vanishing. In the next step, one can construct the total Hamiltonian that generates the temporal evolution of dynamical variables. Since the only primary constraints are the sure constraints \eq{3.2}, the total Hamiltonian is given by
\be
\cH_T=\cH_c+u^i\pi_i{^0}+\frac{1}{2}u^{ij}\Pi_{ij}{^0}\, ,
\ee
where $u^i$ and $u^{ij}$ are canonical multipliers.

By construction, the components $\cH_{ij},\cH_\a$ and $\cH_\orth$ of the
canonical Hamiltonian do not depend on the unphysical variables $b^i{_0}$
and $\om^{ij}{_0}$. Hence, by demanding the primary constraints
to be preserved during the time evolution, one finds the set of
secondary constraints,
\be
\cH_\orth\approx 0\, ,\qquad \cH_\a\approx 0\, ,\qquad \cH_{ij}\approx 0\, . \lab{4.13}
\ee
General arguments, based on the existence of local Poincar\'e invariance,
show that these constraints are first class \cite{x13}, see also
\cite{x28}. Hence, the Dirac consistency algorithm is completed at
the level of the secondary constraints \eq{4.13}.

The present PG model has $N_1=20$ first-class constraints, and $N_2=0$ second-class constraints. Since the number of the Lagrangian variables is $N=40$ (16 tetrad, plus 24 connection components), the number of the Lagrangian degrees of freedom is $N^*=(2N-2N_1-N_2)/2=20$. They are the same as those found in the weak field approximation of PG: 2 massless spin-2 modes and 18 massive torsion modes (two spin-0, six spin-1, and ten spin-2 modes). However, we shall show that not all of these degrees of freedom are physically acceptable, in contrast to earlier expectations \cite{x25}. To do that, we will first calculate the mass eigenvalues $m_\pm^2(J)$ for the torsion modes with spin $J=0,1,2$.

\section{Linearized field equations}\label{sec5}
\setcounter{equation}{0}

In this section, we start our analysis of the particle spectrum of PG by
deriving the weak field approximation of the gravitational field equations
\eq{2.4} around the Minkowski background $M_4$; for consistency, we assume
$\L_0=0$. Such an approximation is based on the following weak field
expansion of the basic dynamical variables,
$$
b^i{_\m}=\d^i_\m+\tilde b^i{_\m}+O_2\, ,\qquad
                       \om^{ij}{_\m}=\tilde\om^{ij}{_\m}+O_2.
$$
To simplify the notation, we omit writing the tilde sign and the symbol
$O_2$, with an implicit understanding of their effects. Furthermore, we
find it technically convenient to use the following abbreviations:
\bea
&&A_n=a_n-a_1\,,\qquad B_n=b_n-b_1\,,                           \nn\\
&&\barA_n=\bara_n-\bara_1\,,\qquad \barB_n=\barb_n-\barb_1\, ,
\eea

\subsection{First field equation}

In the first field equation \eq{2.4a}, the covariant momentum associated
to torsion has the form
\bea
\cH_{imn}&=&2a_1T_{imn}+\frac{2}{3}A_2(\eta_{im}\cV_n-\eta_{in}\cV_m)
                       +2A_3\ve_{imnl}\cA^l\, ,                   \nn\\
  &&-\bara_1 T_{irs}\ve^{rs}{}_{mn}-\frac{2}{3}\barA_2\ve_{imns}\cV^s
     +2\barA_3(\eta_{im}\cA_n-\eta_{in}\cA_m)\, ,
\eea
where $\bara_2=\bara_3$ yields $\bar A_2=\bar A_3$. Then, after calculating the linearized form of $E_i{^\n}$,
$$
E_i{^\n}=2a_0G^\n{_i}
                    -\bara_0\big(R_{mnki}\ve^{mnk\n}+h_i{^\n}X\big)
  =2a_0G^\n{_i}-2\bara_0X_i{^\n}\,,                             \nn
$$
the linearized (1ST) takes the form
\bea
\cE_{in}&=&\pd^m\cH_{imn}
              -2a_0 G_{ni}+2\bara_0X_{in}                       \nn\\
&=&-2a_1\pd^m T_{inm}+\frac{2}{3}A_2(\pd_i\cV_n-\eta_{in}\pd\cV)
                     -2A_3\ve_{inmk}\pd^m\cA^k                  \nn\\
&&+\frac{2}{3}\barA_2\ve_{inmk}\pd^m\cV^k
        +2\barA_2(\pd_i\cA_n-\eta_{in}\pd\cA)-2a_0 G_{ni} +2(\bara_0-\bara_1)X_{in}=0\, ,                         \lab{5.3}
\eea
where we used \eq{E.2}, and $\pd\cV:=\pd_i\cV^i$, $\pd\cA:=\pd_i\cA^i$.

\subsection{Second field equation}

Using the formulas obtained in the weak field approximation,
\bea
\nab_\m {}^LH_{ij}{}^{\m n}
    &=&2a_0(T^n{}_{ij}-\d^n_i\cV_j+\d^n_j\cV_i)                  \nn\\
 &&-\bara_0\ve_{ij}{}^{rs}(T^n{}_{rs}-\d^n_r\cV_s+\d^n_s\cV_r)\,,\nn\\
2\cH_{[ij]n}&=&-\frac{4}{3}(2a_1+a_2)\eta_{n[i}\cV_{j]}
             +2(a_1+2a_3)\ve_{ijnk}\cA^k                         \nn\\
 &&-\frac{4}{3}\barA_2\ve_{ijnk}\cV^k-4\barA_3\eta_{n[i}\cV_{n]}\,,
\eea
the linearized form of (2ND) reads
\bsubeq\lab{5.5}
\bea
\cE_{ijn}&=& \pd^m\cH'_{ijmn}
  +2a_0(T^n{}_{ij}-\d^n_i\cV_j+\d^n_j\cV_i)               \nn\\
 && -\bara_0\ve_{ij}{}^{rs}(T^n{}_{rs}-\d^n_r\cV_s+\d^n_s\cV_r)
    +2\cH_{[ij]n}=0\, .
\eea
Using the double duality relations for the curvature, see Appendix C in Ref. \cite{x29}, the term $\pd^m\cH'_{ijmn}$ is found to have the form
\bea
\pd^m\cH'_{ijmn}&=&
  (b_2+b_1)\pd^m\big(\eta_{ir}\Psi_{js}
                    -\eta_{jr}\Psi_{is}\big)\ve^{rs}{}_{mn}
  +\frac{1}{6}B_3\ve_{ijmn}\pd^m X                                 \nn\\
&&+(b_4+b_1)\Big[\big(\pd_i\Phi_{jn}
                -\eta_{in}\pd^m\Phi_{jm}\big)-(i\lra j)\Big] +\frac{1}{6}B_6(\eta_{jn}\pd_i-\eta_{in}\pd_j)R    \nn\\
&&+B_5\Big[\big(\pd_i\hR_{[jn]}
                -\eta_{in}\pd^m\hR_{[jm]}\big)-(i\lra j)\Big]
  -\barB_5\pd^m\big(\eta_{ir}\hR_{[js]}
                   -\eta_{jr}\hR_{[is]}\big)\ve^{rs}{}_{mn}        \nn\\
&&+(\barb_2-\barb_1)\Big[(\pd_i\Psi_{jn}-\eta_{in}\pd^m\Psi_{jm})
                                              -(i\lra j)\Big]
  +\frac{1}{6}\barB_3(\eta_{jn}\pd_i-\eta_{in}\pd_j)X              \nn\\
&&-(\barb_4-\barb_1)\pd^m\big(\eta_{ir}\Phi_{js}
                             -\eta_{jr}\Phi_{is}\big)\ve^{rs}{}_{mn}
  -\frac{1}{6}\barB_6\ve_{ijmn}\pd^m R\, .                         \lab{5.5b}
\eea
\esubeq

\section{Particle spectrum}\label{sec6}
\setcounter{equation}{0}

The particle spectrum of PG contains important information of its physical
content. Recently, Karananas \cite{x25} made a detailed analysis of this
problem by extending the spin-projection operator formalism, used earlier
in the context of \PG+ \cite{x12}, and applying it to study the PG field
excitations around the Minkowski background. His work resulted in the mass
formulas for  the spin-0, spin-1 and spin-2 massive torsion modes,
together with the related restrictions on the parameter space,  stemming
from the requirements for the absence of ghosts and tachyons.

In this section, we study the same problem by analyzing the linearized
field equations along the lines presented in \cite{x5}. The obtained
results are tested by verifying their compatibility with the expressions
for the critical parameters, found in the canonical analysis, whereas the
absence of ghosts and tachyons is studied in the next section.

\subsection{Spin-0 modes}\label{sub61}

The spin-0 sector is determined by the traces of the field equations
$\cE_{in}$, $\pd^i\cE_{ijn}$ and $\pd^k(\hd\cE)_{kln}$, where
$\hd\cE_{kln}:=(1/2)\ve_{kl}{}^{ij}\cE_{ijn}$ is the dual of $\cE_{ijn}$:
\bea
&&-a_2\pd V-3\bara_2\pd A+a_0 R+\bara_0 X=0\, ,                  \nn\\
&&(b_4+b_6)\square R+(\barb_3-\barb_2)\square X
  +4(2a_0+a_2)\pd V+12(\bara_2-\bara_0)\pd A=0\, ,\qquad         \nn\\
&&(b_2+b_3)\square X-(\barb_3-\barb_2)\square R
  -12(a_0+2a_3)\pd A+8(\bara_2-\bara_0)\pd V=0\, .               \lab{6.1}
\eea
With $X=3\pd A$, the first equation can be used to express $R$ in terms of
$\pd\cV$ and $\pd\cA$, whereupon the remaining two equations are written
in the matrix form as
\bsubeq
\bea\lab{6.2}
&& (K_0\square +4a_0N_0)U=0\, ,                                    \\
&& K_0=\left( \ba{cc}
     a_2(b_4+b_6) &
           -3a_0(\barb_2-\barb_3)-3(\bara_0-\bara_2)(b_4+b_6)      \\
     a_2(\barb_2-\barb_3) &
            3a_0(b_2+b_3)-3(\bara_0-\bara_2)(\barb_2-\barb_3)
          \ea
         \right)\, ,                                               \nn\\
&& N_0=\left( \ba{cc}
         (2a_0+a_2) & -3(\bara_0-\bara_2) \\
       -2(\bara_0-\bara_2) & -3(a_0+2a_3)
            \ea
          \right)\, ,\qquad           U=\left(\ba{c}
                                              \pd\cV \\
                                              \pd\cA
                                              \ea\right)\, .
\eea
The determinants of $K_0$ and $N_0$ are given by
\bea
&&\det K_0=3a_0a_2\big[(b_4+b_6)(b_2+b_3)
                       +(\barb_2-\barb_3)^2\big]\, ,               \nn\\
&&\det N_0=-3\big[(2a_0+a_2)(a_0+2a_3)+2(\bara_0-\bara_2)^2\big]\, .
\eea
\esubeq
For $\det K_0\ne 0$, one can multiply \eq{6.2} by $K_0^{-1}$, and obtain the Klein-Gordon equation for the massive spin-0 torsion modes,
\be
(\square +M_0)U=0\, ,\qquad  M_0=4a_0K_0^{-1}N_0\, .              \lab{6.3}
\ee
The masses of these modes are given by the eigenvalues of the mass matrix $M_0$,
\bsubeq\lab{6.4}
\bea\lab{6.4a}
m^2_\pm(0)&=&\frac{1}{2}\Big(\tr M_0\pm\sqrt{(\tr M_0)^2-4(\det M_0)}\,\Big)\nn\\
 &=&\frac{2a_0}{\det K_0}\Big(\tr f_0+\sqrt{(\tr f_0)^2-4\,\det f_0}\,\Big)\,,
\eea
where $f_0:=(\det K_0)K_0^{-1}N_0$, and
\bea
\tr f_0&=&3a_0(2a_0+a_2)(b_2+b_3)
          -12a_0(\bara_0-\bara_2)(\barb_2-\barb_3)               \nn\\
     &&-3\big[a_2(a_0+2a_3)+2(\bara_0-\bara_2)^2\big](b_4+b_6)\,,\nn\\
\det f_0&=&(\det K_0)(\det N_0)\, .
\eea
\esubeq
It is interesting to note that $\det K_0$ is proportional
to the product of two critical parameters, $a_2$ and $\det B_0$, characterizing the spin-$0$ sector of the set of if-constraints (see Table 1). Hence, when the critical parameters vanish, we have $\det K_0=0$, the mass eigenvalues \eq{6.4} become infinite, and consequently, the spin-0 modes do not propagate. In the linear regime, this mechanism provides a Lagrangian description of the dynamical role of if-constraints.

As a further test of our mass formula \eq{6.4}, we calculated its form in the parity-even sector $(\bara_0,\bara_n,\barb_n)=0$, and found the well-known result for the spin-$0^\pm$ torsion modes:
$$
m_+^2(0)=\frac{4a_0(2a_0+a_2)}{a_2(b_4+b_6)}\, ,\qquad
m_-^2(0)=-\frac{4(a_0+2a_3)}{(b_2+b_3)}\, .
$$

\subsection{Spin-1 modes}\label{sub62}

To understand the linearized dynamics of the spin-1 sector, it is convenient to start with the antisymmetric part of (1ST), $\cE_{[ij]}$, and its dual, $\hd\cE_{ij}$. Taking derivatives of these equations yields
\bea
&&\frac{1}{3}(2a_1+a_2)(\square\cV_j-\pd_j\pd\cV)
     +\barA_3(\square\cA_j-\pd_j\pd\cA)
     +2A_0\pd^i\hR_{[ij]}+2\barA_0\pd^i X_{[ij]}=0\,,           \nn\\
&&(a_1+2a_3)(\square\cA_j-\pd_j\pd\cA)
     -\frac{2}{3}\barA_2(\square\cV_j-\pd_j\pd\cV)
     -2A_0\pd^iX_{[ij]}+2\barA_0\pd^i\hR_{[ij]}=0\, .           \nn
\eea
Then, the solutions for $\pd^m\hR_{[mi]}$ and $\pd^m X_{[mi]}$
are found to be given in the matrix form as
\bea\lab{6.5}
&&2\left(\ba{c}
       -\pd^m\hR_{[mi]} \\
        \pd^m X_{[mi]}
       \ea\right)=G\left(\square U_i-\pd_i(\pd U)\right)\, ,\qquad
  U_i=\left(\ba{c}
           \cV_i \\
           \cA_i
           \ea\right)\, ,\qquad  g:=A_0^2+\barA_0^2\, ,         \nn\\
&&G:=\frac{1}{g}\left(\ba{cc}
      \dis\frac{1}{3}\big[A_0(2a_1+a_2)-2\barA_0\barA_2\big]  &
                     \Big[A_0\barA_2+\barA_0(a_1+2a_3)\Big]   \\
      -\dis\frac{1}{3}\big[\barA_0(2a_1+a_2)+2A_0\barA_2\big] &
                     -\big[\barA_0\barA_2-A_0(a_1+2a_3)\big]
           \ea\right)\,.
\eea

Next, consider the trace of (2ND), $\eta^{jk}\cE_{ijk}$, and of its dual, $\eta^{jk}\hd\cE_{ijk}$. Using the identities \eq{E.3}, these trace components take the form
\bsubeq\lab{6.6}
\bea
&&-2(b_4+b_5)\pd^m\hR_{[mi]}+2(\barb_2-\barb_5)\pd^m X_{[mi]}
  +\frac{1}{2}(b_4+b_6)\pd_i R
         -\frac{1}{2}(\barb_2-\barb_3)\pd_i X                   \nn\\
&&\ph{xxx} +2(2a_0+a_2)\cV_i-6(\bara_0-\bara_2)\cA_i=0\,,       \\
&&-4(\barb_2-\barb_5)\pd^m\hR_{[mi]}-4(b_2+b_5)\pd^m X_{[mi]}
         +(\barb_2-\barb_3)\pd_i R+(b_2+b_3)\pd_i X             \nn\\
&&\ph{xxx} -8(\bara_0-\bara_2)\cV_i-12(a_0+2a_3)\cA_i=0\,.
\eea
\esubeq
Using the expressions for $\pd^m\hR_{[mi]}$ and $\pd^mX_{[mi]}$ found in \eq{6.5}, and the expression for $R$ determined by the trace of (1ST), Eqs. \eq{6.6} multiplied by $-2g$ can be written in the matrix form as
\bsubeq
\be
\big(K_1\square-4gN_1\big)U_i+(L_1-K_1)\,\pd_i(\pd U)=0\,,      \lab{6.7a}
\ee
where
\bea
&&K_1=B'_1(gG)\, ,\qquad B'_1:=-2\left(\ba{cc}
                                b_4+b_5 & \barb_2-\barb_5 \\
                           2(\barb_2-\barb_5) & -2(b_2+b_5)
                                \ea\right)  \, ,\qquad
                                       \nn\\
&&N_1=\left( \ba{cc}
          (2a_0+a_2)   & -3(\bara_0-\bara_2) \\
          -4(\bara_0-\bara_2) & -6(a_0+2a_3)
         \ea\right)\, , \qquad U_i=\left(\ba{c}
                                         \cV_i \\
                                         \cA_i
                                         \ea\right)\, ,         \nn\\
&&L_1=-\frac{g}{a_0}\left(\ba{cc}
                         1 & 0 \\
                         0 & 2
                        \ea\right)K_0=-4gN_1M_0^{-1}\, .
\eea
The determinants of $K_1$ and $N_1$ are given by
\bea
&&\det K_1=\frac{8}{3}g(\det A)(\det B_1)\,,                    \nn\\
&&\det N_1=-6\Big[(a_0+2a_3)(2a_0+a_2)+2(\bara_0-\bara_2)^2\Big]\,.\lab{6.7c}
\eea
\esubeq
When $\det K_1\ne 0$, one can multiply Eq. \eq{6.7a} by $K_1^{-1}$ and obtain the matrix Klein-Gordon equation for the massive spin-1 torsion modes,
\bea
&&(\square +M_1)\tU_i=0\,,\qquad M_1:=-4gK_1^{-1}N_1\, ,        \nn\\
&&\tilde U_i:=U_i+M_0^{-1}\pd_iU\, ,\qquad \pd^i\tU_i=0\, .     \lab{6.8}
\eea

The eigenvalues of the mass matrix $M_1$ are given by
\bsubeq\lab{6.9}
\be
m_\pm^2(1)=\frac{-2g}{\det K_1}
      \Big(\tr f_1\pm\sqrt{(\tr f_1)^2-4\det f_1}\,\Big)\, ,
\ee
where $f_1:=(\det K_1)K_1^{-1}N_1$, and
\bea
&&\det f_1=(\det N_1)(\det K_1)\, ,                             \nn\\
&&\tr f_1=4(b_2+b_5)\Big[
        (2a_0+a_2)\big[(a_0-a_1)(a_1+2a_3)-(\bara_0-\bara_1)^2\big]
                         +2(a_0-a_1)(\bara_0-\bara_2)^2 \Big]   \nn\\
  &&\qquad+4(b_4+b_5)\Big[(a_0+2a_3)\big[(a_0-a_1)(2a_1+a_2)
                          -2(\bara_0-\bara_1)^2\big]
                          +2(a_0-a_1)(\bara_0-\bara_2)^2\Big]   \nn\\
  &&\qquad +8(\barb_2-\barb_5)\Big[
           -(2a_0+a_2)(a_0+2a_3)(\bara_0-\bara_1)+2\big[(a_0-a_1)^2
            +(\bara_0-\bara_1)^2\big](\bara_0-\bara_2)          \nn\\
  &&\qquad -2(\bara_0-\bara_1)(\bara_0-\bara_2)^2\Big]\, .
\eea
\esubeq

The determinant of $K_1$ is the product of two critical parameters associated to the spin-1 sector (see Table 1). A discussion of what happens when at least one of these parameters vanishes is given in Appendix \ref{appD}.

In the parity even sector, our mass formula \eq{6.9} yields
the familiar result for the spin-1$^\pm$ torsion modes:
$$
m_+^2(1)=\frac{6(a_0-a_1)(a_0+2a_3)}{(a_1+2a_3)(b_2+b_5)}\, ,\qquad
m_-^2(1)=\frac{6(a_0-a_1)(2a_0+a_2)}{(2a_1+a_2)(b_4+b_5)}\,.
$$

\subsection{Spin-2 modes}\label{sub63}

Although, in principle, the analysis of the spin-2 sector is not much more complicated than the one for the spin-1 case, the fact that there are lots of variables makes the general procedure rather complex and difficult to follow. In Ref. \cite{x27}, the mass eigenvalues of the spin-2 torsion modes
were found by studying a class of exact wave solutions, defined by an ansatz that creates only the tensorial irreducible part of the torsion, whereas the vector and axial vector parts vanish. This motivates us to simplify the present discussion by considering a dynamical system with vanishing spin-0 and spin-1 modes, $\cV_i=0$ and $\cA_i=0$. The physical content of such a system is described solely by the spin-2 tensor $t_{ijk}$ (Appendix \ref{appA}). Such a technical simplification does not influence the final result for the spin-2 mass eigenvalues.

The adopted assumptions have two additional consequences: $X=0$, which follows from $X=3\pd\cA$, and $R=0$, which follows from the trace of (1ST).
To analyse the spin-2 sector, we need the symmetrized version of (1ST),
\be
-a_1\Th_{ik}-a_0\Phi_{ik}+\barA_0\Psi_{ik}=0\, ,                \lab{6.10}
\ee
where $\Th_{ik}:=\pd^m t_{ikm}=\pd^m T_{(ik)m}$, as follows from the definition \eq{A.1} of $t_{ikm}$. Moreover, we also need two equations that follow from (2ND), $\pd^m\cE_{m(ik)}$, and $\pd^m(\hd\cE)_{m(ik)}$:
\bsubeq\lab{6.11}
\bea
&&(b_1+b_4)\big[\square\Phi_{ik}-2\pd_{(i}\pd^m\Phi_{k)m}\big]
  +\barB_2\big[\square\Psi_{ik}-2\pd_{(i}\pd^m\Psi_{k)m}\big]   \nn\\
&&\ph{xxx}-2A_0\Th_{ik}-2\barA_0\Psi_{ik}=0\, ,                 \\[3pt]
&&(b_1+b_2)\big[\square\Psi_{ik}-2\pd_{(i}\pd^m\Psi_{k)m}\big]
  -\barB_4\big[\square\Phi_{ik}-2\pd_{(i}\pd^m\Phi_{k)m}\big]   \nn\\
&&\ph{xxx}-2\barA_0\Th_{ik}+2A_0\Psi_{ik}=0\,.
\eea
\esubeq
Since $\Phi_{ik}$ has a nontrivial Riemannian part associated to the massless graviton, a proper description of the torsion spin-2 modes is obtained by using \eq{6.10} to eliminate $\Phi_{ik}$ from Eqs. \eq{6.11}:
\bsubeq
\bea
&&(b_1+b_4)\square\big(-a_1\Th_{ik}+\barA_0\Psi_{ik}\big)
  +a_0\barB_2\square\Psi_{ik}
  -2a_0\big(A_0\Th_{ik}+\barA_0\Psi_{ik}\big)=0\, ,            \\[3pt]
&&a_0(b_1+b_2)\square\Psi_{ik}
  -\barB_2\square\big(-a_1\Th_{ik}+\barA_0\Psi_{ik}\big)
  -2a_0\big(\barA_0\Th_{ik}-A_0\Psi_{ik}\big)=0\,.\qquad
\eea
\esubeq
These equations can be compactly represented in the matrix form as
\be
(K_2\square+2a_0 N_2)U_{ik}=0\,,                                \lab{6.13}
\ee
where
\bea
&&K_2:=\left(\ba{cc}
     a_1(b_1+b_4)      & -\barA_0(b_1+b_4)-a_0(\barb_2-\barb_1) \\
   -a_1(\barb_2-\barb_1) & \barA_0(\barb_2-\barb_1)-a_0(b_2+b_1)
       \ea\right)\, ,                                           \nn\\
&&N_2:=\left(\ba{cc}
         A_0 & \barA_0 \\
         \barA_0 & -A_0
          \ea\right)\, ,\qquad
  U_{ik}:=\left(\ba{c}
                \Th_{ik} \\
                \Psi_{ik}
                \ea\right)\, .                                  \nn
\eea
For $\det K_2\ne 0$, Eq. \eq{6.13} is equivalent to
\be
\big(\square+M_2)U_{ik}=0\, ,\qquad M_2:=2a_0K_2^{-1}N_2\, ,
\ee
where $M_2$ is the mass matrix of the spin-2 torsion mode.

The matrices $K_2$ and $N_2$ are of the same form as those found in Ref. \cite{x27}, Eq. (4.50), up to inessential differences in conventions. Hence, the mass eigenvalues are also the same. Expressed in terms of the matrix $f_2=(\det K_2)K_2^{-1}N_2$, they are given by
\bsubeq\lab{6.15}
\be
m^2_\pm(2)=\frac{a_0}{\det K_2}
           \Big(\tr f_2\pm\sqrt{(\tr f_2)^2-4\det f_2}\,\Big)\,,\lab{6.15a}
\ee
where
\bea
\det f_2&=&(\det K_2)(\det N_2)\, ,                             \nn\\
\tr f_2&=&-a_0(a_0-a_1)(b_1+b_2)
          -2a_0(\bara_0-\bara_1)(\barb_2-\barb_1)               \nn\\
 && +\big[-a_1(a_0-a_1)+(\bara_0-\bara_1)^2\big](b_1+b_4)\, .   \lab{6.15b}
\eea
\esubeq
As expected, the determinant of $K_2$ is proportional to the product of the critical parameters given in the third line of Table 1,
\be
\det K_2=-a_0a_1\det B_2\,,\qquad
\det N_2=-\big(A_0^2+\barA_0^2\big)\, .
\ee

In the parity-even sector, the above formulas produce the well-known result:
$$
m^2_+(2)=\frac{2a_0(a_0-a_1)}{a_1(b_1+b_4)}\, ,\qquad
m^2_-(2)=\frac{2(a_0-a_1)}{b_1+b_2}\, .
$$

The above procedure can be extended to the case with nonvanishing spin-0 and spin-1 terms. After a straightforward but rather clumsy calculation, we found that the new terms do not influence the mass eigenvalues, they only modify the spin-2 state $U_{ik}$. A compact form of the result reads
\bsubeq
\be
U_{ik}\to\tU_{ik}:=Z\left[{\bar U_{ik}}-G\pd_{(i}U_{k)}
     +\frac{1}{3}H\Big(M_0^{-1}\pd_i\pd_k U+\eta_{ik}U\Big)\right]\,,
\ee
where
\bea
&&Z:=\frac{1}{a_1}\left(\ba{cc}
                 -a_0 & \barA_0 \\
                   0  &  a_1
                        \ea\right)\,,\qquad
      {\bar U_{ik}}:=\left(\ba{c}
                           \Phi_{ik} \\
                           \Psi_{ik}
                           \ea\right)\, ,                       \nn\\
&& H:=\frac{1}{a_0}\left(\ba{cc}
                       a_2 & 3(\bara_2-\bara_0) \\
                       0   & -2a_0
                         \ea\right)\, .
\eea
\esubeq
The role of $Z$ is to replace $\Phi_{ik}$ in $\bar U_{ik}$ by its form obtained from the symmetrized (1ST). The spin-2 nature of $\tU_{ik}$ is ensured by the properties $\pd^i\tU_{ik}=0$, $\eta^{ik}\tU_{ik}=0$. In fact, these properties are sufficient to completely determine $\tU_{ik}$.

\subsection{Comparison with Karananas' mass formulas}

Our mass formulas are found to be consistent with the expressions for the
canonical critical parameters, displayed in Table 1. A more detailed test
can be conducted by comparing them to the recent calculations of Karananas \cite{x25}. The first step in this direction is to compare the
Lagrangian (5) in Ref. \cite{x25} with our expression \eq{B.1}. Although the
procedure is straightforward, a number of misprints found in \cite{x25}
complicate the process. Nevertheless, we  established the following correspondence between the related parameters:
\be
\ba{ll}
a_0=\l\, ,                     & \bara_0=\L=0\, ,              \\
a_1=\l+t_1\, ,                 & a_2=2(-\l+t_3)\, ,            \\
a_3=(-\l+t_2)/2\,,             &                               \\
\bara_1=-2t_5\,,               & \bara_2=\bara_3=-t_4\,,       \\
b_1=4(r_1-r_3)\,,              & b_2=4r_3\,,                   \\
b_3=4(r_2-r_3)\, ,             & b_4=4(r_1-r_3+r_4)\,,         \\
b_5=4(r_3+r_5)\,,              & b_6=4(r_1-r_3+3r_4)\, ,       \\
\barb_1=r_7-3r_8\,,            & \barb_2=\barb_4=r_7+r_8\,,    \\
\barb_3=\barb_6=-4r_6+r_7+r_8,\quad &\barb_5=-3r_7+r_8\, .     \lab{6.18}
\ea
\ee
The remaining part of the comparison is rather simple. By substituting the
above expressions into Eqs. \eq{6.4} and \eq{6.9}, one finds that the
resulting mass eigenvalues for the spin-0 and spin-1 torsion modes exactly
reproduce the respective result that Karananas gives in his Appendix A.
Moreover, we also found that, up to minor differences, our mass formula \eq{6.15} for the spin-2 modes is in agreement with his result (A.3.5); see also subsection IV.E in Ref. \cite{x27}. Although the difference is small, it might be responsible for more serious discrepancies in the physical properties of the spin-2 modes, found in the next section.

\section{Physical restrictions on the space of parameters}\label{sec7}
\setcounter{equation}{0}

In this section, we study the physical requirements of the absence of ghosts ($E>0$), the absence of tachyons ($m^2>0$), and the reality ($m^2$ real), in the spectrum of torsion modes. Our approach is based on the Hamiltonian analysis developed in Secs. \ref{sec3} and \ref{sec4}, subject to the assumption that all the critical parameters are nonvanishing, or equivalently,
that all the torsion modes are propagating. In what follows, we shall examine whether such an assumption is compatible with the adopted physical requirements.

Our general strategy is the following. The conditions of the \emph{positivity of energy} can be read from the dynamical component $\cH^R_\orth$ of the canonical Hamiltonian, see Eq. \eq{4.11}.
By introducing the matrices
$$
F_J:=\frac{1}{\det B_J}R_J\, ,\qquad J=0,1,2,
$$
these conditions can be expressed by demanding that the eigenvalues of $F_j$ be positive. Using the general formula for the eigenvalues of a $2\times 2$ matrix, see \eq{3.9}, one can express these conditions in a more practical form as
\be
E_J>0:\qquad \det  F_J>0\, ,\quad \tr F_j>0\, .                 \lab{7.1}
\ee
The absence of tachyons is effectively described by the conditions of \emph{positivity} of the eigenvalues $m_\pm^2(J)$ of the mass matrices $M_J$:
\be
m^2_\pm(J)>0:\qquad\det M_J>0\, ,\quad \tr M_J>0\, .            \lab{7.2}
\ee
Moreover, the presence of square roots in the mass eigenvalues requires to check their \emph{reality}:
\be
m^2_\pm(J)~\text{real:}\qquad (\tr M_J)^2-4\det M_J>0\, .       \lab{7.3}
\ee

By applying these general physical criteria to the specific spin-$J$ sectors, one obtains a set of restrictions on the original Lagrangian parameters.
An important goal of our analysis is to clarify the issue of their mutual (in)consistency. We shall always use $a_0>0$, the condition that ensures the correct limit to GR.

\subsection{Spin-0 sector}

\subsubsection*{Positivity of energy}

The energy of the spin-0 modes is positive if the eigenvalues of the
matrix $F_0=R_0/\det B_0$ are positive. Since $\det R_0=4\det B_0$, the first condition $\det F_0>0$ implies that $\det B_0>0$,  or equivalently,
\bsubeq\lab{7.4}
\bea
&&(b_2+b_3)(b_4+b_6)+(\barb_2-\barb_3)^2<0\, ,                  \lab{7.4a}\\
&&\Ra\quad (b_2+b_3)(b_4+b_6)<0\, .                             \lab{7.4b}
\eea
\esubeq
Then, the second condition takes the form $\tr R_0>0$. In combination with \eq{7.4b}, it yields the relations
\be
b_2+b_3<0\, ,\qquad b_4+b_6>0\, ,                               \lab{7.5}
\ee
which coincide with those in appearing in \PG+. The independent conditions are the condition \eq{7.4a} and, for instance, the first one in \eq{7.5}
\be
(b_2+b_3)(b_4+b_6)+(\barb_2-\barb_3)^2<0\,,
                                      \qquad b_2+b_3<0\,.       \lab{7.6}
\ee
These two conditions coincide with the first two relations found in Eq.
(48) of Ref. \cite{x25} (the third relation is redundant).

\subsubsection*{Positivity of \mb{m_\pm^2(0)}}

The mass matrix $M_0$ of the spin-0 torsion modes has the form \eq{6.3},
\be
M_0/4a_0=K_0^{-1}N_0=\frac{1}{\det K_0}f_0\, ,\qquad
  \det K_0=-3a_0a_2\det B_0\, .
\ee
The positivity of its eigenvalues is expressed by the conditions $\det M_0>0$ and $\tr M_0>0$:
\be
\frac{\det N_0}{\det K_0}>0\, ,\qquad \frac{1}{\det K_0}\tr f_0>0\, .
\ee
Since $\det B_0>0$, they take the form
\bsubeq\lab{7.9}
\bea
&&a_2\det N_0<0\,,                                               \lab{7.9a}\\
&&a_2\tr f_0<0\, .                                               \lab{7.9b}
\eea
\esubeq

As shown in Appendix \ref{appF}, these general conditions can be transformed into an unexpectedly simple form, in which the parameters $(b_n\barb_n)$ are completely absent:
\be
a_2\left[(2a_0+a_2)(a_0+2a_3)+2(\bara_0-\bara_2)^2\right]>0\, ,
\qquad a_2(2a_0+a_2)>0\, .
\ee
Returning to the parameters introduced in \eq{6.18}, this result takes the form
$$
(t_3-\l)(t_2t_3+t_4^2)>0\,,\qquad (t_3-\l)t_3>0\,.
$$
The first formula is equivalent to Karananas' result \cite{x25}, but the second one is different.

\subsection{Spin-1 sector}

\subsubsection*{Positivity of energy}

Starting with $F_1:=R_1/\det B_1$ and using $\det R_1=-\det B_1$, the first condition for the positivity of energy, $\det F_1>0$, reads
\bsubeq\lab{7.11}
\bea
&&\det B_1\equiv(b_2+b_5)(b_4+b_5)+(\barb_2-\barb_5)^2<0\,,     \lab{7.11a}\\
&&\Ra\quad (b_2+b_5)(b_4+b_5)<0\, .                             \lab{7.11b}
\eea
\esubeq
The second condition, written as $\tr R_1<0$ and combined with \eq{7.11b}, yields
\be
b_2+b_5>0\, ,\qquad b_4+b_5<0\,,                                \lab{7.12}
\ee
which is the \PG+ result. As the two independent conditions, we choose
\be\lab{7.13}
(b_2+b_5)(b_4+b_5)+(\barb_2-\barb_5)^2<0\,,\quad~b_4+b_5<0\,.
\ee
Again, there is a complete agreement with the first two relations in Eq. (49) of \cite{x25}, whereas the third relation is redundant.

\subsubsection*{Positivity of \mb{m_\pm^2(1)}}

To make the technical exposition more compact, we introduce the following notation:
\bea
&&\m_2:=2a_0+a_2\, ,\quad \m_3:=a_0/2+a_3\, ,
  \quad k_2:=2a_1+a_2\, ,\quad k_3:=a_1/2+a_3\,.\quad             \nn\\
&&\det A=-2\Big[k_2k_3+(\bara_1-\bara_2)^2\Big]\,,\qquad
  \det N_1=-12\Big[\m_2\m_3+(\bara_0-\bara_2)^2\Big]\,.         \nn
\eea
The mass matrix of the spin-1 torsion modes was found in subsection \ref{sub62},
\be
M_1=-4gK_1^{-1}N_1=-\frac{4g}{\det K_1}f_1\, ,\qquad
 \det K_1=\frac{8}{3}g(\det A)(\det B_1)\, ,
\ee
with $g\equiv A_0^2+\barA_0^2$.
The positivity of the mass eigenvalues is expressed by the requirements
\be
\frac{\det N_1}{\det K_1}>0\, ,\qquad \frac{1}{\det K_1}\tr f_1<0\, .
\ee
Since $\det B_1<0$, these conditions are equivalent to
\bsubeq\lab{7.16}
\bea
&&(\det A)(\det N_1)<0\, ,                                    \lab{7.16a}\\
&&(\det A)\tr f_1>0\, .                                       \lab{7.16b}
\eea
\esubeq
The expression for $\tr f_1$ is given in subsection \ref{sub62}; see also Appendix \ref{appF}.

A simple inspection of \eq{7.16a} shows that it can be realized by $\det A<0,\det N_1>0$, or vice versa, whereas, as shown in Appendix \ref{appF}, \eq{7.16b} can be replaced by a much simpler expression. The resulting conditions, equivalent to \eq{7.16}, are defined in \eq{F.7}:
\bea
\text{(i)}&&k_2k_3+(\bara_1-\bara_2)^2<0\,,\qquad
             \mu_2\mu_3+(\bara_0-\bara_2)^2>0\,,                \nn\\
       &&\mu_3k_2A_0-2\m_3\bar A_0^2+A_0(\bara_0-\bara_2)^2<0\,,\nn\\
\text{(ii)}&&k_2k_3+(\bara_1-\bara_2)^2>0\,,\qquad
           \mu_2\mu_3+(\bara_0-\bara_2)^2<0\,,                  \nn\\
       &&\mu_3k_2A_0-2\m_3\bar A_0^2+A_0(\bara_0-\bara_2)^2>0\,.
\eea
As before, they do not depend on the parameters $(b_n,\barb_n)$.
Going over to the parameters defined in \eq{6.18}, the relations (i) read
\bea
&&(t_1+t_2)(t_1+t_3)+(t_4-2t_5)^2<0\,,\qquad t_2t_3+t_4^2>0\,,  \nn\\
&&t_2(t_1^2+4t_5^2)+t_1(t_2t_3+t_4^2)>0\,.                      \nn
\eea
The first two inequalities in the set (i) coincide with those found in Ref. \cite{x25}, the third one is a bit different, but the whole complementary set (ii) is missing.

\subsection{Spin-2 sector}\label{sec73}

\subsubsection*{Positivity of energy}

The first condition for the positivity of the eigenvalues of $F_2=R_2/\det B_2$, $\det F_2>0$, combined with $\det R_2=-4\det B_2$, takes the form
\bsubeq
\bea
&&\det B_2\equiv(b_1+b_2)(b_1+b_4)+(\barb_2-\barb_1)^2<0\, ,   \lab{7.18a}\\
&&\Ra\quad (b_1+b_2)(b_1+b_4)<0\, .                            \lab{7.18b}
\eea
\esubeq
The second condition combined with \eq{7.18b} yields relations
that are also valid in \PG+,
\be
b_1+b_2<0\, ,\qquad b_1+b_4>0\, .                              \lab{7.19}
\ee
The two independent conditions are
\be
(b_1+b_2)(b_1+b_4)+(\barb_2-\barb_1)^2<0\,,
                          \qquad~b_1+b_2<0\,.                  \lab{7.20}
\ee
Comparing these conditions to the first two relations in Eq. (50) of Ref. \cite{x25}, one finds a complete agreement (the third relation is redundant).

\subsubsection*{Positivity of \mb{m_\pm^2(2)}}

The mass matrix for the spin-2 modes is found in subsection \ref{sub63}:
\bea
&&M_2=2a_0K_2^{-1}N_2=\frac{2a_0}{\det K_2}f_2\, ,\qquad
                      \det K_2=-a_0a_1\det B_2\,,               \nn\\
&&\det N_2=-\big[(a_0-a_1)^2+(\bara_0-\bara_1)^2\big]\,.        \nn
\eea
The positivity of the mass eigenvalues is expressed by the requirements
\be
\frac{\det N_2}{\det K_2}>0\, ,\qquad
\frac{2a_0}{\det K_2}\tr f_2>0\, .
\ee
The condition $\det N_2<0$ implies $\det K_2<0$, whereupon, relying on $\det B_2<0$, one obtains
\bsubeq\lab{7.22}
\bea
&& a_1<0\, ,                                               \lab{7.22a}\\
&&\tr f_2<0\, ,                                            \lab{7.22b}
\eea
\esubeq
where $\tr f_2$ is calculated in subsection \ref{sub63}.

\subsubsection*{Is the spin-2 sector free of ghosts and tachyons?}

Let us recall that in \PG+, the conditions $a_1<0$ and $b_1+b_2<0$ imply $\tr f_2>0$, so that one of the two spin-2$^\pm$ modes is always a tachyon, as is well known. In what follows, we will prove, somewhat unexpectedly, that the same conclusion also holds in the general PG.

To show this, we rewrite $\tr f_2$ in a compact notation as
\bsubeq
\be
\tr f_2=\a_2(b_1+b_2)+\b_2(\barb_2-\barb_1)+\g_2(b_1+b_4)\, ,\qquad\a_2<0\, ,
\ee
where the coefficients $\a_2,\b_2$ and $\g_2$ can be read from Eq. \eq{6.15b},
\bea
&&\a_2=-a_0(a_0-a_1)\, ,\qquad \b_2=-2a_0(\bara_0-\bara_1)^2\, ,\nn\\
&&\g_2=-a_1(a_0-a_1)+(\bara_0-\bara_2)^2\, ,                    \nn
\eea
and $\a_2<$ follows from \eq{7.22a}. Since $b_1+b_4>0$, one finds
\be
\frac{\tr f_2}{b_1+b_4}=\a_2\frac{b_1+b_2}{b_1+b_4}
              +\b_2\frac{\barb_2-\barb_1}{b_1+b_4}+\g_2\, .     \lab{7.23b}
\ee
Having in mind the first relation in \eq{7.20}, written as
$$
\frac{b_1+b_2}{b_1+b_4}+x^2<0\, ,\qquad
                          x:=\frac{\barb_2-\barb_1}{b_1+b_4}\,,
$$
we find it useful to rewrite \eq{7.23b} in an equivalent form,
\bea
&&\frac{\tr f_2}{b_1+b_4}=
  \a_2\left(\frac{b_1+b_2}{b_1+b_4}+x^2\right)+F_2(x)\, ,       \nn\\
&&F_2(x):=-\a_2x^2+\b_2x+\g_2\, .                              \lab{7.23c}
\eea
\esubeq
A critical argument in our analysis comes from the observation that the discriminant of the quadratic function $F_2(x)$, $\D_2=\b_2^2+4\a_2\g_2$, is automatically negative,
\be
\D_2=4a_0a_1\Big[(\bara_0-\bara_1)^2+(a_0-a_1)^2\Big]
                 =-4a_0a_1\det N_2<0\,.                         \lab{7.24}
\ee
Next, since $\a_2<0$ (the parabola $F_2$ opens upward) and $\D_2/\a_2>0$ (minimum of $F_2$ is positive), it follows that $F_2(x)>0$ for any $x$. Hence,
using \eq{7.20}, one obtains the result
\be
\tr f_2>(b_1+b_4)F_2(x)>0\, ,                                   \lab{7.25}
\ee
which contradicts to \eq{7.22b}. Thus
\bitem
\item[{\bf S3.}] The two no-tachyon conditions, \eq{7.22a} and \eq{7.22b}, are mutually exclusive; hence, the two spin-2 torsion modes cannot be  simultaneously physical.
\eitem
Such a conclusion is not in agreement with the result found by Karananas \cite{x25}.

\subsubsection*{No-ghost conditions: spin-2 versus spin-1 sector}

The no-ghost conditions for spin-1 and spin-2 sectors are in contradiction to each other. Indeed, \eq{7.12} implies that $b_2>b_4$, whereas \eq{7.19} implies that $b_4>b_2$. Hence, only one of these two sectors can be physical. The result is in agreement with the Corrigendum in \cite{x25}.

\subsection{Reality conditions}

The structure of the general reality conditions \eq{7.3} looks rather cumbersome. However, after replacing $|\tr f_0|$, $|\tr f_1|$, and $|\tr f_2|$ with their minimal values, calculated from the inequalities \eq{F.3}, \eq{F.6}, and \eq{7.25}, respectively, the reality conditions \eq{7.3} transform into
\bea
\text{spin 0:}
&&(b_4+b_6)^2a_2\det N_0+12a_0(2a_0+a_2)^2\det B_0<0\, ,        \nn\\
\text{spin 1:}
&&g(b_2+b_5)^2(\det A)(\det N_1)-24\a_1^2\det B_1<0\, ,         \nn\\
\text{spin 2:}
&&(b_1+b_4)^2a_1\det N_2+4a_0(a_0-a_1)^2\det B_2>0\, ,          \lab{7.26}
\eea
see Sec. \ref{sec6} and Appendix \ref{appF}. These formulas are much simpler than \eq{7.3}, but they represent only sufficient conditions for the reality of the corresponding mass eigenvalues.

\section{Summary and conclusions}\label{sec8}
\setcounter{equation}{0}

In this paper, we investigated generic aspects of the Hamiltonian structure of the general parity-violating PG, and used them to study the torsion particle spectrum \cite{x30}.

Making use of Dirac's Hamiltonian approach, we identified the set of all \emph{if-constraints}, the expressions that become true constraints if the corresponding \emph{critical parameters} $c_n$ vanish. Both the if-constraints and the associated critical parameters have a crucial influence on the PG dynamics. Then, we constructed the generic form of the \emph{canonical Hamiltonian} $\cH_c$, determined by taking all the critical parameters to be nonvanishing. An extension of the procedure to allow for a proper treatment of the vanishing critical parameters is outlined in Appendix \ref{appD}.

Apart from being important by itself, the Hamiltonian structure introduced here turns out to be intrinsically related to the particle spectrum of PG.
To examine that subject, we first calculated the \emph{mass eigenvalues}  $m^2_\pm(J)$ of the torsion modes with spin $J=0,1$, and $2$, relying on the weak field approximation of the gravitational field equations around $M_4$. As a test of the results, we verified that $m_\pm^2(J)$ are proportional to the inverse critical parameters $1/c_n$. As a consequence, whenever some of $c_n$ vanish, the corresponding  values of $m^2_\pm(J)$ become infinite, thereby preventing the associated torsion modes to propagate. This is consistent with the canonical effects of the vanishing critical parameters in \PG+ (in the weak field approximation). A comparison of our mass formulas to those found by Karananas \cite{x25} leads to the following conclusions.
\bitem
\item[(k1)] For the spin-0 and spin-1 torsion modes, our results confirm those of Karananas.\vspace{-9pt}
\item[(k2)] For the spin-2 modes, there are certain differences, noted already in Ref. \cite{x27}.
\eitem

The \emph{absence of ghosts} (positivity of energy) in the particle spectrum is ensured by demanding the positivity of the specific spin-$J$ terms in the canonical Hamiltonian, whereas the conditions for the \emph{absence of tachyons} are defined by the requirement $m^2_\pm(J)>0$. A detailed analysis shows that these requirements can be formulated as follows:
$$
\ba{rl}
\text{Spin 0:}&
    (b_2+b_3)(b_4+b_6)+(\barb_2-\barb_3)^2<0\,,~~~ b_2+b_3<0\,, \\[3pt]
  &a_2\left[(2a_0+a_2)(a_0/2+a_3)+(\bara_0-\bara_2)^2\right]<0\,,
    \qquad a_2(2a_0+a_2)>0.                                     \\[9pt]
\text{Spin 1:}&
  (b_2+b_5)(b_4+b_5)+(\barb_2-\barb_5)^2<0\,,\quad~b_4+b_5<0\,, \\[3pt]
\text{(i)}& (2a_1+a_2)(a_1/2+a_3)+(\bara_1-\bara_2)^2<0,\quad
             (2a_0+a_2)(a_0/2+a_3)+(\bara_0-\bara_2)^2>0,       \\[3pt]
       & (a_0-a_1)\big[(a_0/2+a_3)(2a_1+a_2)+(\bara_0-\bara_2)^2\big]
         -2(a_0/2+a_3)(\bara_0-\bara_1)^2<0\,;                  \\[3pt]
\text{(ii)}& \text{an alternative set of conditions, obtained by
                      (i) $\to$ $(-1)\times$(i).}               \\[9pt]
\text{Spin 2:}&(b_1+b_2)(b_1+b_4)+(\barb_2-\barb_1)^2<0\, ,\qquad
               b_1+b_2<0\, ,                                    \\[2pt]
 & \text{the conditions for the absence of tachyons are \emph{mutually
         exclusive}.}                                       \hfill (8.2)
\ea
$$
The results for the absence of ghosts (first line in each spin sector) are identical to those of Karananas, whereas the formulas describing the absence of tachyons show a number of less or more serious differences. In particular, the whole set of conditions (ii) in the spin-1 sector is missing in Karananas' analysis, but the most important difference is found in the spin-2 sector, where the two conditions for the absence of tachyons are in contradiction to each other, in contrast to Karananas' conclusion.

The presence of square roots in the expressions for the mass eigenvalues $m^2_\pm(J)$ requires to verify their reality. A sufficient form of of the \emph{reality conditions}, compactly presented at the end of Sec. \ref{sec7}, is much simpler than their general form.

In conclusion, our analysis clarifies the structure of the particle spectrum of the general PG by improving the results found by Karananas, in particular the status of the spin-2 sector. On the other hand, elements of the Hamiltonian structure introduced here, including its extension to the case of vanishing critical parameters outlined in Appendix \ref{appD}, are a good starting point for further investigation of the full nonlinear dynamics of PG.

\section*{Acknowledgments}

We are very grateful to Georgios Karananas for his readiness to discuss all
aspects of his earlier work, and for a careful reading of the manuscript.
We would also like to thank Friedrich Hehl for useful comments. This work was partially supported by the Serbian Science Foundation under Grant No. 171031.

\appendix
\section{Irreducible decomposition of the field strengths}\label{appA}
\setcounter{equation}{0}

The torsion tensor has three irreducible pieces:
\bsubeq\lab{A.1}
\bea
&&{}^{(2)}T_{imn}=\frac{1}{3}(\eta_{im}\cV_n-\eta_{in}\cV_m)\,, \nn\\
&&{}^{(3)}T_{imn}=\ve_{imnk}\cA^k\, ,                           \nn\\
&&{}^{(1)}T_{imn}=T_{imn}-{}^{(2)}T_{imn}-{}^{(3)}T_{imn}
                 =\frac{4}{3}t_{i[mn]}\, ,
\eea
where
\bea
&&\cV_n:=T^k{}_{kn}\,,\qquad \cA_k:=\frac{1}{6}\ve_{krst}T^{rst}\,,\nn\\
&&t_{imn}:=T_{(im)n}+\frac{1}{3}\eta_{n(i}\cV_{m)}
                  -\frac{1}{3}\eta_{im}\cV_n\,.
\eea
\esubeq
The Riemann-Cartan curvature tensor can be decomposed into six irreducible pieces:
\bsubeq
\bea
&&{}^{(2)}R_{ijmn}
  =\frac{1}{2}(\eta_{ik}\Psi_{jl}-\eta_{jk}\Psi_{il})
                                        \ve{}^{kl}{}_{mn}\, ,   \nn\\
&&{}^{(3)}R_{ijmn}=\frac{1}{12}X\ve_{ijmn}\, ,                  \nn\\
&&{}^{(4)}R_{ijmn}=\frac{1}{2}(\eta_{im}\Phi_{jn}-\eta_{jm}\Phi_{in})
                                               -(m\lra n)\,,    \nn\\
&&{}^{(5)}R^{ij}=\frac{1}{2}(\eta_{im}\hR_{[jn]}-\eta_{jm}\hR_{[in]})
                                               -(m\lra n)\,,    \nn\\
&&{}^{(6)}R_{ijmn}=\dis\frac{1}{12}R(\eta_{im}\eta_{jn}
                                    -\eta_{jm}\eta_{im})\, ,    \nn\\
&&{}^{(1)}R_{ijmn}=R_{ijmn}-\sum_{a=2}^6{}^{(a)}R_{ijmn}\, ,   \lab{A.2a}
\eea
where
\be
\ba{ll}
\hR_{im}:=\ric_{im}=R_{inm}{^n}\, ,\qquad &R:=\ric^m{_m}\,,     \\[3pt]
X_{ij}:=\dis\frac{1}{2}R_{ikmn}\ve^{kmn}{_j}\,,\qquad &X:=X^n{_n}\,,\\[3pt]
\Phi_{ij}:=\ric_{(ij)}-\dis\frac{1}{4}\eta_{ij}R\,,\qquad
  &\Psi_{ij}:=X_{(ij)}-\dis\frac{1}{4}\eta_{ij}X\, .
\ea                                                             \lab{A.2b}
\ee
\esubeq

The above definitions are the tensor counterparts of the corresponding formulas given in terms of the differential forms, see \cite{x27,x29}. They imply the following relations characterizing the parity-odd sector:
\bsubeq\lab{A.3}
\bea
&&T^{ijk}\,\hd\irr{2}{T}_{ijk}=T^{ijk}\,\hd\irr{3}{T}_{ijk}\, ,     \nn\\
&&R^{ijkl}\,\hd\irr{2}{R}_{ijkl}=R^{ijkl}\,\hd\irr{4}{R}_{ijkl}\, , \nn\\
&&R^{ijkl}\,\hd\irr{3}{R}_{ijkl}=R^{ijkl}\,\hd\irr{6}{R}_{ijkl}\, , \lab{A.3a}
\eea
and also
\bea
&&T^{ijk}\,\hd\irr{1}{T}_{ijk}
  =\irr{1}{T}^{ijk}\,\hd\irr{1}{T}_{ijk}\, ,                      \nn\\
&&R^{ijkl}\hd\,\irr{1}{R}_{ijkl}
  =\irr{1}{R}^{ijkl}\,\hd\irr{1}{R}_{ijkl}\, ,                    \nn\\
&&R^{ijkl}\,\hd\irr{5}{R}_{ijkl}
  =\irr{5}{R}^{ijkl}\,\hd\irr{5}{R}_{ijkl}\, .
\eea
\esubeq

\section{Alternative form of the Lagrangian}\label{appB}
\setcounter{equation}{0}

In this appendix, we rewrite our Lagrangian \eq{2.6} in an equivalent form
that allows an easier comparison to the literature \cite{x22,x25}:
\bsubeq\lab{B.1}
\bea
\cL_G&=&-(a_0R+2\L_0+\bara_0X)+\cL_{T^2}+\cL_{R^2}\,,           \nn\\[3pt]
\cL_{T^2}&=&h_1T^{ijk}T_{ijk}+h_2T^{imn}T_{nmi}+h_3V_{m}V^n     \nn\\[3pt]
    &&+\ve^{mnkl}(\bh_4T^i{}_{mn}T_{ikl}
                      +\bh_5T_{mn}{^i}T_{kli})\, ,              \\
\cL_{R^2}&=&\frac{1}{2}\big(f_1R^{ijmn}R_{ijmn}+f_2R^{ijmn}R_{imjn}
          +f_3R^{ijmn}R_{mnij}                                  \nn\\
  &&+f_4\ric^{im}\ric_{im}+f_5\ric^{im}\ric_{mi}+f_6R^2\big)    \nn\\
  &&+\frac{1}{2}\ve^{mnkl}\big(\barf_7R_{mnkl} R
            +\barf_8R_{ijmn}R^{ij}{}_{kl}                       \nn\\
  &&+\barf_9R_{mnij}R_{kl}{}^{ij}
            +\barf_{10}R_{mnij}R^{ij}{}_{kl}\big)\, .
\eea
\esubeq
The parameters $(h_n,\bh_n)$ and $(f_n,\barf_n)$ can be expressed in terms
of the ``irreducible" parameters appearing in \eq{2.6} as follows:
\bsubeq
\bea
&&h_1=\frac{1}{6}(2a_1+a_3)\, ,\qquad h_2=\frac{1}{3}(a_1-a_3)\,,
  \qquad  h_3=-\frac{1}{3}(a_1-a_2)\,,                          \nn\\
&&\bh_4:=-\frac{1}{24}(4\bara_1+\bara_2+\bara_3)\, ,\qquad
  \bh_5:=-\frac{1}{6}(2\bara_1-\bara_2-\bara_3)\, ,
\eea
and
\bea
&&f_1:=\frac{1}{12}(2b_1+3b_2+b_3)\,,\qquad
   f_2:=\frac{1}{3}(b_1-b_3)\, ,                                \nn\\
&&f_3:=\frac{1}{12}(2b_1-3b_2+b_3)\,,\qquad
         f_4:=-\frac{1}{2}(b_1+b_2-b_4-b_5)\,,                  \nn\\
&&f_5:=-\frac{1}{2}(b_1-b_2-b_4+b_5)\, ,\qquad
         f_6:=\frac{1}{12}(2b_1-3b_4+b_6)\, ,                   \nn\\
&&\barf_7:=\frac{1}{24}(2\barb_1-\barb_3-\barb_6)\, ,\qquad
  \barf_8:=-\frac{1}{16}(\barb_1+\barb_2+\barb_4+\barb_5)\,,~   \nn\\
&&\barf_9:=-\frac{1}{16}(\barb_1-\barb_2-\barb_4+\barb_5)\,,\qquad
  \barf_{10}:=-\frac{1}{8}(\barb_1-\barb_5)\, .
\eea
\esubeq

Relying on the existence of three topological invariants \eq{2.9}, Karananas
\cite{x25} imposed three conditions on the Lagrangian parameters in \eq{B.1}:~
$\bara_0,f_6,\barf_8=0$.

Based on \eq{B.1}, one can find a suitable form of the covariant momenta $\cH_{ijk}$ and $\cH'_{ijkl}$:
\bsubeq\lab{B.3}
\bea
\cH_{ijk}&=&4\big(h_1T_{ijk}+h_2T_{[kj]i}
                              +h_3\eta_{i[j}V_{k]}\big)\, ,     \nn\\
    &&+4\big(\bh_4\ve_{jk}{}^{mn}T_{imn}
                   +\bh_5\ve_{i[j}{}^{mn}T_{mnk]}\big)\,,       \\
\cH'_{ijkl}&=&4\Big[ f_1R_{ijkl}
             +\frac{1}{2}f_2\big(R_{i[kjl]}-R_{j[kil]}\big)
             +f_3R_{klij}\Big]                                  \nn\\
  &&+2\bigl[
     f_4\big(\eta_{j[l}\hR_{ik]}-\eta_{i[l}\hR_{jk]}\big)
    +f_5\big(\eta_{j[l}\hR_{k]i}-\eta_{i[l}\hR_{k]j}\big)
    +f_6\big(\eta_{jl}\eta_{ik}-\eta_{il}\eta_{jk}\big)R\big]   \nn\\
  &&+\barf_7\big[ 2\ve_{ijkl}R
       + (\ve R)(\eta_{ik}\eta_{jl}-\eta_{jk}\eta_{il}\big)\big]\nn\\
  &&+4\big(
    \barf_8\ve_{kl}{}^{mn}R_{ijmn}+\barf_9\ve_{ij}{}^{mn}R_{mnkl}\big)
    +2\barf_{10}\big(\ve_{ij}{}^{mn}R_{klmn}
                 +\ve_{kl}{}^{mn}R_{mnij}\big). \qquad
\eea
\esubeq
\section{\mb{(3+1)} decomposition of spacetime}\label{appC}
\setcounter{equation}{0}

The dynamical content of canonical constraints is greatly clarified by
using a decomposition of tensor fields with respect to the subgroup of
3d rotations in the spatial hypersurface $\S_0:~ x^0=$ const.

Let \mb{e}$_\a$ be a basis of three coordinate tangent vectors in $\S_0$, \mb{e}$_\a=\pd_\a~ (\a=1,2,3)$, and \mb{n} the unit normal to $\S_0$, $n_k=h_k{^0}/\sqrt{g^{00}}$. The four vectors (\mb{n},\mb{e}$_\a$) define the so-called ADM basis of tangent vectors in spacetime.
The decomposition of the vector \mb{e}$_0$ in the ADM basis is given by
\bsubeq\lab{C.1}
\be
\mb{e}_0=N\mb{n}+N^\a\mb{e}_\a\, ,
\ee
where $N$ and $N^\a$, known as the lapse and shift functions,
respectively,  are linear in $b^k{_0}$:
\bea
&&N=\mb{e}_0\mb{n}=n_k b^k{_0}=1/\sqrt{g^{00}}\, ,              \nn\\
&&N^\a=\mb{e}_0\mb{e}_\b\ir{3}g^{\b\a}=h_\bk{^\a}b^k{_0}=-g^{0\a}/g^{00}\, .
\eea
\esubeq

Introducing the projectors on \mb{n} and \mb{e}$_\a$, given by
$(P_\orth)_k^l=n_kn^l$ and $(P_{||})_k^l=\d_k^l-n_kn^l$, respectively,
one can express a spacetime vector $V_k$ in terms of its orthogonal (to
$\S_0$) and ``parallel" (living in the tangent space of $\S_0$)
components:
\be
V_k=n_k V_\orth+V_\bk\, ,                                       \lab{C.2}
\ee
where $V_\orth:=n^k V_k$ and $V_\bk:=V_k-n_k V_\orth$ Here, we use a
convention that a bar over an index `k' denotes its parallel projection,
so that $n^kV_\bk$ vanishes. The objects $V_\orth$ and $V_\bk$ are
respectively a scalar and a vector with respect to 3d rotations in $\S_0$.

Consider now a second-rank tensor $X_{ik}$. Its orthogonal-parallel
decomposition reads
\be
X_{ik}=n_iX_{\orth\bk}+n_in_k X_{\orth\orth}+n_kX_{\bi\orth}+X_{\bi\bk}\, .
\ee
Here, $X_{\orth\bk}$ is a vector and $X_{\orth\orth}$ a scalar with
respect to 3d rotations, whereas the irreducible parts of $X_{\bi\bk}$ are
its trace, antisymmetric and traceless symmetric parts:
\bsubeq\lab{C.4}
\bea
&&\ir{S}X:=X^\bk{_\bk}\, ,\quad\ir{A}X_{\bi\bk}:=X_{[\bi\bk]}\, ,\quad
  \ir{T}X_{\bi\bk}:=X_{(\bi\bk)}-\frac{1}{3}\eta_{\bi\bk}X^\bm{_\bm}\,,\nn\\
&&X_{\bi\bk}=\ir{A}X_{\bi\bk}+\ir{T}X_{\bi\bk}
                             +\frac{1}{3}\eta_{\bi\bk}\ir{S}X\, .
\eea
As a consequence,
\be
X^{\bi\bk}Y_{\bi\bk}=\ir{A}X_{\bi\bk}\ir{A}Y_{\bi\bk}
    +\ir{T}X_{\bi\bk}\ir{T}Y_{\bi\bk}+\frac{1}{3}\ir{S}X\ir{S}Y\, .
\ee
\esubeq

Now, it is straightforward to extend these considerations to any tensor.
As a particularly interesting example, we consider the spacetime tensor
$X_{ijk}=-X_{jik}$, which is decomposed into the set of spatial tensors
$(X_{\orth\bj\orth},X_{\orth\bj\bk},X_{\bi\bj\orth},X_{\bi\bj\bk})$. The
irreducible parts of $X_{\bi\bj\bk}=-X_{\bj\bi\bk}$ are the pseudoscalar, the vector and the traceless symmetric tensor, respectively:
\bsubeq\lab{C.5}
\bea
&&\ir{P}X:=\ve^{\bi\bj\bk}X_{\bi\bj\bk}\, ,\qquad
  \ir{V}X_\bi:=X_{\bi\bk}{}^\bk\, ,                             \nn\\
&&\ir{T}X_{\bi\bj\bk}:=X_{\bi(\bj\bk)}
      +\frac{1}{2}\eta_{\bi(\bj}\,\ir{V}X_{\bk)}
       -\frac{1}{2}\eta_{\bj\bk}\,\ir{V}X_\bi\, .
\eea
The tensor part satisfies the cyclic identity
$\ir{T}X_{\bi\bj\bk}+\ir{T}X_{\bk\bi\bj}+\ir{T}X_{\bj\bk\bi}=0$.
The epsilon tensor $\ve_{\bi\bj\bk}$ is defined by
$\ve_{\bi\bj\bk}:=\ve_{\orth\bi\bj\bk}$ and satisfies the identities
\bea
&&\ve_{\bi\bj\bk}\ve^{\bm\bn\bk}
          =-(\d^\bm_\bi\d^\bn_\bj-\d^\bm_\bj\d^\bn_\bi)\, ,     \nn\\
&&\ve_{\bi\bn\bk}\ve^{\bm\bn\bk}=-3\d^\bm_\bi\,,
  \qquad \ve_{\bm\bn\bk}\ve^{\bm\bn\bk}=-6\, .                  \nn
\eea
The related decomposition formulas read:
\bea
&&X_{\bi\bj\bk}=\frac{4}{3}\ir{T}X_{[\bi\bj]\bk}
                     -\eta_{\bk[\bi}\ir{V}X_{\bj]}
                          -\frac{1}{6}\ve_{\bi\bj\bk}\ir{P}X\,, \nn\\
&&X^{\bi\bj\bk}Y_{\bi\bj\bk}=
    \frac{4}{3}\ir{T}X^{\bi\bj\bk}\ir{T}Y_{\bi\bj\bk}
         +\ir{V}X^\bi\ir{V}Y_\bi-\frac{1}{6}\ir{P}X\ir{P}Y\, .
\eea
\esubeq

\section{General construction of \mb{\cH_\orth}}\label{appD}
\setcounter{equation}{0}

In this appendix, we discuss the general structure of $\cH_\orth$, including  the case when some of the critical parameters vanish. In a simplified but self-evident notation, the relations that define critical parameters have the following typical form (see Sec. \ref{sec3}):
\be
\vphi=F V\, ,                                                   \lab{D.1}
\ee
where
$$
\vphi:=\left(\ba{c}
              \vphi_1 \\
              \vphi_2
              \ea\right)\, ,\qquad
 F:=\left(\ba{cc}
          a & \barb \\
          \barc & d
          \ea\right)\, , \qquad V:=\left(\ba{c}
                                        V_1 \\
                                        V_2
                                        \ea\right)\, .
$$
Here, $\vphi$ represents the if-constraints, $V$ are the corresponding velocities, and $F$ is the matrix with eigenvalues $c_1,c_2$. Since $F$ is chosen to represent $A,B_0,B_1$ or $B_2$, the parameter $\barc$ is proportional to $\barb$, $\barc=\k\barb$. If $\barb=0$, the matrix $F$ is already diagonal, and the construction of $\cH_\orth$ is quite simple. When $\barb\ne 0$, which is typical for the parity-violating PG, the matrix $F$ needs first to be diagonalized. The diagonal form $D$ of $F$ is constructed as
\be
D=P^{-1}FP\,,\qquad P:=\left(\ba{cc}
                          -\barb & -\barb \\
                           a-c_1 & a-c_2
                             \ea\right)\,,\qquad
D=\left(\ba{cc}
        c_1  & 0  \\
         0   & c_2
        \ea\right)\, ,                                          \lab{D.2}
\ee
where $P$ is invertible provided $\det P=\barb(c_2-c_1)\ne 0$, and
$$
P^{-1}=\frac{1}{\det P}\left(\ba{cc}
                             a-c_2 & \barb \\
                            -a+c_1 & -\barb
                            \ea\right)\, .
$$
Left multiplication of \eq{D.1} by $P^{-1}$ yields
\bsubeq\lab{D.3}
\be
\vphi'=DV'\, ,
\ee
where $\vphi':=P^{-1}\vphi$ and $V':=P^{-1}V$, or equivalently,
\be
\vphi'_1=c_1 V'_1\, ,\qquad \vphi'_2=c_2 V'_2\, .               \lab{D.3b}
\ee
\esubeq

To construct the related $F$-part of $\cH_\orth$, note that its typical form reads
\be
\cH_\orth^F=\vphi^TQV\equiv\vphi'{}^T(P^TQP)V'\, ,\qquad
Q:=\left(\ba{cc}
         q_1 & 0 \\
          0  & q_2
          \ea\right)\, ,                                        \lab{D.4}
\ee
see Sec. \ref{sec4}. Further discussion depends on the specific values of $c_1$ and $c_2$.

(1) When $c_1,c_2\ne 0$, Eq. \eq{D.1} implies $V=F^{-1}\vphi$, and  $\cH^F_\orth=\vphi^TQF^{-1}\vphi$ coincides with the result found in Sec. \ref{sec4}.
(2) The case $c_1=c_2=0$ is rather trivial: both if-constraints $\vphi'_n$ become true constraints that appear in the total Hamiltonian, but $\cH^F_\orth=0$.
(3) Finally, when only one critical parameter vanishes (which requires $\det F=0$), say $c_2=0$, then $\vphi'_2=0$ (a new constraint), $V'_2$ remains undetermined and $\vphi'_1=c_1V'_1$. Hence, \eq{D.4} implies that
\be
\cH^F_\orth=(\barb^2q_1+d^2q_2)\frac{1}{c_1}(\vphi'_1)^2
            +\vphi'_1(\barb^2q_1-adq_2)V'_2\, .                 \lab{D.5}
\ee
The result can be also expressed in terms of the original if-constraints $\vphi_n$ by noting that $\vphi'_2=0$ implies $\vphi'_1=-\vphi_1/\barb$. The factor $1/c_1$ in the first term shows a typical dependence on the critical parameters, known from \PG+, whereas the second term, linear in the undetermined velocity $V_2'$,  can be absorbed into the total Hamiltonian, see \cite{x14,x15,x23}. The presence of an extra constraint $\vphi'_2$ requires to complete the whole consistency procedure.

In the context of the weak field approximation, the form of $\cH^F_\orth$ in \eq{D.5} determines the no-ghost conditions for the case (3):
\be
\det F=ad-\barb\barc=0\, ,\qquad \s c_1>0\, ,                   \lab{D.6}
\ee
where $\s$ is the sign of $(\barb^2q_1+d^2q_2)$ and $c_1=a+d$.

Now, we have a comment on kind of ``non-analiticity" of the above results. Since the assumption $\barb\ne 0$ ensures the regularity of the matrix $P$, the diagonal matrix $D$ in \eq{D.2} has no valid limit for $\barb\to 0$. Hence, the expressions for $c_n$ when $\barb=0$ cannot be obtained by taking the limit $\barb\to 0$ of the generic result. However, since the matrix $F$ for $\barb=0$ is already diagonal, the critical parameters $c_n$ can be obtained directly from $F$. The same conclusion also holds for the form of $\cH^F_\orth$.

\section{Linearized Bianchi identities}\label{appE}
\setcounter{equation}{0}

In Secs. \ref{sec5} and \ref{sec6}, many technical simplifications were
obtained with the help of the linearized Bianchi identities,
\be
\ve^{\m\n\l\r}R^{ij}{}_{\n\l\r}=0\, ,\qquad
\ve^{\m\n\l\r}\pd_\n T_{i\l\r}=\ve^{\m\n\l\r}R_{i\n\l\r}\, ,
\ee
and their consequences. In particular, the first identity implies that
\be
\pd^k X_{ik}=0\,,\qquad \pd^i G_{ik}=0\, ,
\ee
where $G_{ik}:=\ric_{ik}-(1/2)\eta_{ik}R$, and the second identity yields
\bea
&&X_i{}^j=-\frac{1}{2}\ve^{jkmn}\pd_k T_{imn},\qquad X=3\pd\cA\,,\nn\\
&&\ve^{ijmn}R_{ijmk}=2X_k{^n}-\d_k^nX\, ,                       \nn\\
&&2\ric_{[mn]}=-\pd^k T_{kmn}+2\pd_{[m}\cV_{n]}\, .             \lab{E.2}
\eea
As a consequence,
\be
\pd^m\Phi_{im}=\pd^m\hR_{[im]}+\frac{1}{4}\pd_i R\,,\qquad
  \pd^m\Psi_{im}=\pd^m X_{[mi]}-\frac{1}{4}\pd_i X\, .         \lab{E.3}
\ee

\section{Simplified conditions for the absence of tachyons}\label{appF}
\setcounter{equation}{0}

In this appendix, we derive a simplified form of the conditions \eq{7.9} and \eq{7.16}, describing the absence of tachyons in the spin-0 and spin-1 sectors, respectively; the spin-2 sector is discussed in subsection \ref{sec73}.

\subsubsection*{Spin-0 sector}

The expression for $\tr f_0$, found in subsection \ref{sub61}, can be represented in a suitable form as
\bsubeq
\be
\frac{1}{3}\tr f_0=\a_0(b_2+b_3)+\b_0(\barb_2-\barb_3)
                                +\g_0(b_4+b_6)\, ,              \lab{F.1a}\\
\ee
where
\bea
&&\a_0=a_0(2a_0+a_2)\,,\qquad \b_0=4a_0(\bara_2-\bara_0)\,,     \nn\\  &&\g_0=-\big[a_2(a_0+2a_3)+2(\bara_0-\bara_2)^2\big]\, .        \nn
\eea
After dividing this equation by $(b_4+b_6)>0$, one obtains
$$
\frac{\tr f_0}{3(b_4+b_6)}=
  \a_0\frac{b_2+b_3}{b_4+b_6}+\b_0\frac{\barb_2-\barb_3}{b_4+b_6}+\g_0\, .
$$
By noting that the first relation in \eq{7.6} can be written as
$$
\frac{b_2+b_3}{b_4+b_6}+x^2<0\, ,\qquad x:=\frac{\barb_2-\barb_3}{b_4+b_6}\,,
$$
we find it useful to rewrite \eq{F.1a} in the form
\bea
&&\frac{\tr f_0}{3(b_4+b_6)}=
  \a_0\left(\frac{b_2+b_3}{b_4+b_6}+x^2\right)+F_0(x)\, ,       \lab{F.1b}\\
&&F_0(x):=-\a_0x^2+\b_0x+\g_0\, .                               \nn
\eea
\esubeq
Further analysis is based on an important property of the quadratic function $F_0(x)$, based on \eq{7.9a}: its discriminant, $\D_0=\b_0^2+4\a_0\g_0$, is always negative,
\be
\D_0=-4a_0a_2\left[(2a_0+a_2)(a_0+2a_3)+2(\bara_0-\bara_2)^2\right]
 \equiv(4/3)a_0a_2\det N_0<0\, .                                \lab{F.2}
\ee

Similar considerations applied to $a_2\tr f_0$ modify Eq. \eq{F.1b} by an overall multiplicative factor $a_2$. To simplify the discussion, we introduce  a suitable notation: $\a_0':=a_2\a_0$ and $F'_0(x):=a_2F_0(x)$. Note that the discriminant $\D_0'$ of the new function $F'_0(x)$ remains negative. Now, we are ready to prove the following statement:
\bitem
\item[{\bf S0.}] Given $\D_0<0$, the condition $\a_0'\equiv a_2\a_0>0$ is equivalent to $a_2\tr f_0<0$.
\eitem

To prove this equivalence, we start by assuming $\a'_0>0$, which implies
\be
a_2\tr f_0<3(b_4+b_6)F'_0(x)\, .                                \lab{F.3}
\ee
Moreover, the parabola $F'_0(x)$ opens downward, and $\D'_0/\a'_0<0$ (negative at vertex) ensures that $F'_0(x)<0$ for any $x$. Hence, $a_2\tr f_0<0$, what was to be shown.

The reverse statement $a_2\tr f_0<0\Ra\a'_0>0$  can be easily proven by reductio ad absurdum, that is by showing that $\a_0'<0$ implies $a_2\tr f_0>0$, which is a contradiction.

The statement {\bf S0} allows us to replace \eq{7.9b} with the much simpler condition $a_2>0$.

\subsubsection*{Spin-1 sector}

For the spin-1 sector, we first rewrite $\tr f_1$ in the form
\bsubeq
\be
\frac{1}{4}\tr f_1=
           \a_1(b_4+b_5)+\b_1(\barb_2-\barb_5)+\g_1(b_2+b_5)\,, \lab{F.4a}
\ee
where
\bea
&&\a_1:=2\mu_3k_2A_0-4\m_3\bar A_0^2+2A_0(\bara_0-\bara_2)^2\,, \nn\\
&&\b_1:=-4\m_2\m_3\barA_0+4(A_0^2+\barA_0^2)(\bara_0-\bara_2)
       -4\barA_0(\bara_0-\bara_2)^2\, ,                         \nn\\
&&\g_1:=2A_0\m_2k_3-\m_2\barA_0^2+2A_0(\bara_0-\bara_2)^2\,.    \nn
\eea
After dividing by $(b_2+b_5)>0$, one can rewrite \eq{F.4a} in a suitable form
\bea
&&\frac{1}{4(b_2+b_5)}\tr f_1
  =\a_1\left(\frac{b_4+b_5}{b_2+b_5}+x^2\right)+F_1(x)\, ,         \lab{F.4b}\\
&&F_1(x):=-\a_1x^2+\b_1x+\g_1\,,\qquad x:=\frac{\barb_2-\barb_5}{b_2+b_5}\,,\nn
\eea
\esubeq
As a consequence of \eq{7.16a}, the discriminant $\D_1$ of the quadratic function $F_1(x)$ is automatically negative,
\bea
\D_1&:=&16(A_0^2+\barA_0^2)\big[\m_2\mu_3+(\bara_0-\bara_2)^2\big]
       \big[k_2k_3+(\bara_1-\bara_2)^2\big]                       \nn\\
    &&\equiv\frac{2}{3}(A_0^2+\barA_0^2)(\det N_1)(\det A)<0\, .
\eea

To relate our considerations to the properties of $(\det A)\tr f_1$, we multiply Eq. \eq{F.4b} by $\det A$, and introduce a suitable notation $\a_1':=(\det A)\a_1$ and $F_1'(x):=(\det A)F_1(x)$. The new discriminant $\D_1'$ is also negative. Now, one can prove the following statement:
\bitem
\item[{\bf S1.}] For $\D_1<0$, the condition $\a'_1\equiv (\det A)\a_1<0$ is equivalent to $(\det A)\tr f_1>0$.
\eitem

The proof goes as follows. Starting with $\a'_1<0$, one obtains
\be
(\det A){\rm tr}f_1>4(b_2+b_5)F'_1(x)\, .                        \lab{F.6}
\ee
Then, by noting that the parabola opens upward ($\a_1'<0$) and $\D'_1/\a_1'>0$ (positive at vertex), one concludes that $F'_1(x)>0$. Hence, $(\det A)\tr f_1>0$.

As before, the reverse statement $(\det A)\tr f_1>0\Ra\a'_1<0$ can be proven by showing that $\a'_1>0$ leads to $(\det A)\tr f_1<0$, which is a contradiction.

The condition $\D_1<0$, combined with $(\det A)\a_1<0$, can be realized in two ways:
\bea
\text{(i)}&&\det A>0\, ,\qquad \det N_1<0\, ,\qquad  \a_1<0\, . \nn\\
\text{(ii)}&&\det A<0\, ,\qquad \det N_1>0\, ,\qquad \a_1>0\, . \lab{F.7}
\eea



\begin{thebibliography}{99}
\bibitem{x30} Calculations performed in this paper are checked using the Excalc package of the computer algebra system Reduce, and Wolfram Mathematica.
\end{thebibliography}
\end{document}